\pdfoutput=1
\documentclass{ws-rv961x669}
\usepackage{ws-rv-van}            % numbered citation/references
\usepackage{ws-rv-thm}  % comment when other thm package is used
\newcommand\spm{\mathrel{\text{\framebox[0.9\width]{$\pm$}}}}
\newcommand\smp{\mathrel{\text{\framebox[0.9\width]{$\mp$}}}}

\newcommand\splus{\boxplus}%%{\fbox{{$+$}}}
\newcommand\sminus{\boxminus}%%{\fbox{{$-$}}}
\usepackage{subfigure}
\makeindex
\begin{document}

\chapter[Spin-charge-family theory]{The Spin-Charge-Family theory offers the explanation for all 
the assumptions of the Standard model, for the Dark matter,  for the Matter-antimatter 
asymmetry, making several predictions}

\author[N.S.Manko\v c Bor\v stnik]{Norma Susana Manko\v c Bor\v stnik}
%\aindx{Author, F.}                          % author index entry
\address{University of Ljubljana, FMF, Department of Physics, Jadranska 19, 1000 Ljubljana,
Slovenia}
\begin{abstract}
The  {\it spin-charge-family} theory~\cite{JMP2015,norma2014MatterAntimatter,JMP,%
NBled2013,NBled2012,norma92,norma93,norma94,pikanorma,portoroz03,norma95,%
gmdn07,gn,gn2013,gn2015,NPLB,N2014scalarprop}, which is a kind of the Kaluza-Klein 
theories but with 
fermions carrying two kinds of spins (no charges), offers the explanation for all the assumptions
of the {\it standard model}, with the origin of families, the higgs and the Yukawa couplings 
included.
It offers the explanation also for other phenomena, like  the origin of the dark matter and of the 
matter/antimatter asymmetry in the universe.
It predicts the existence of the fourth family to the observed three, as well as several scalar fields
with the weak and the hyper charge of the {\it standard model} higgs ($\pm \frac{1}{2},
\mp \frac{1}{2}$, respectively), which determine the mass matrices of family members, 
offering an explanation, why the fourth family with the masses above $1$ TeV contributes 
weakly to the gluon-fusion production of the observed higgs and to its decay into two 
photons~\cite{AhmedAliMatthiasNeubert},  and predicting that the two photons events, 
observed at the LHC at $\approx 750$ GeV~\cite{twophotonsCMS,twophotonsATLAS,%
twophotonsLHC}, might be an indication for the existence of one of several scalars predicted 
by this theory. 
\end{abstract}

\body

\section{Introduction}
\label{introduction}

The {\it spin-charge-family} theory~\cite{JMP2015,norma2014MatterAntimatter,JMP,NBled2013,%
NBled2012,norma92,norma93,norma94,pikanorma,portoroz03,norma95,gmdn07,gn,gn2013,%
gn2015,NPLB,N2014scalarprop} offers the explanation for all the assumptions of the 
{\it standard model}: {\bf i.$\;$} For the properties of each family member - quarks and leptons,
left and right handed (right handed neutrinos are in this theory regular members of each family) 
and for anti-fermions.  {\bf ii.$\;$} For the appearance of the families.  {\bf iii.$\;$} For the 
existence of the gauge vector fields of the family members charges.  {\bf iv.$\;$} For the scalar 
field and the Yukawa couplings. It is offering the explanation also for the phenomena, which are 
not included in the {\it standard model}, like:  {\bf v.$\;$} 
For the existence of the dark matter~\cite{gn}.  {\bf vi.$\;$} For the (ordinary) 
matter-antimatter asymmetry~\cite{norma2014MatterAntimatter} in the universe.

The {\it spin-charge-family} theory  predicts that there are at the low energy regime two 
decoupled groups of four families: The fourth~\cite{NBled2013,JMP,pikanorma,portoroz03,gmdn07}
to the 
already observed three families of quarks and leptons will be measured at the LHC~\cite{gn2013}. 
At the LHC observed two photons event at $\approx 750$ GeV~\cite{twophotonsCMS,%
twophotonsATLAS,twophotonsLHC}  might be due to the scalar field, which mostly couples to the
$4^{th}$ family quarks~\cite{norma2016higgs}. The lowest of the upper four families 
builds the dark matter~\cite{gn}.

The $4 \times 4$ mass matrices of all the family members demonstrate in this theory the same 
symmetry~\cite{gn2013,gn2015}, determined by the scalar fields: The two $\widetilde{SU}(2)$ 
triplets - the gauge fields of the two family groups operating among families - and the three 
singlets - the gauge fields of the three charges, ($Q, Q'$ and $Y'$), distinguishing among family
members~\cite{norma2014MatterAntimatter,JMP2015}. All these scalar fields carry the weak 
and the hyper charge as does the scalar of the {\it standard model}: ($\pm \frac{1}{2}$ and 
$ \mp \frac{1}{2}$, respectively)~\cite{N2014scalarprop}. 

Since there has been no direct observation of the fourth family quarks with the masses below $1$
TeV, while the fourth family quarks with masses above $1$ TeV would contribute according to the 
{\it standard model} to either the
gluon-fusion scalar field (the higgs) production or to the scalar field decay to two photons 
$\approx 10$ times too much in comparison with the observations~\footnote{ According to the 
{\it standard model} there are  the Yukawa couplings which determine the couplings of quarks 
to the scalar higgs, making them proportional to $\frac{m^{\alpha}_{i}}{v}$, if $m^{\alpha}_{i}$
is the $i^{th}$ family member ($\alpha=u,d$) mass and $v$ the vacuum expectation value of 
the scalar.}, the high energy physicists do not expect the existence of the fourth family members 
at all~\cite{AhmedAliMatthiasNeubert}. 

Might this mean that there does not exist the fourth family coupled to the observed three?
Let be pointed out again that  the {\it spin-charge-family} theory  is able - while starting from a very 
simple action in $d \ge (13+1)$, Eqs.~(\ref{action}, \ref{faction}), with massless fermions with 
the spin of two kinds (one kind taking care of the spin and the charges of the family members,
the second kind taking care of the families~\footnote{The two kinds of spins are connected with 
the left and the right multiplication of any Clifford algebra object~\cite{JMP2015}, respectively.}),
which couple only to the gravity (through the vielbeins and the two kinds of the corresponding 
spin connections) - to explain not only all the assumptions of the {\it standard model}: 
{\bf a.}~It explains the appearance of all the charges of the left and the right handed members.
{\bf b.}~It explains the appearance of all the corresponding vector and scalar gauge fields and 
their properties (explaining the higgs and the Yukawa couplings). 
{\bf c.}~It explains the appearance and properties of family members and their families.
It  answers also several open questions beyond the {\it standard model}, like:
{\bf d.}~It explains the appearance and properties of the dark matter~\cite{gn}. 
{\bf e.}~It explains the appearance of the matter/antimatter asymmetry in the
universe~\cite{norma2014MatterAntimatter}, as well as the proton decay.
{\it The more work is done on the} {\it spin-charge-family} {\it theory, the more explanations of
the observed phenomena and the more predictions for the future observations follow.}

This paper  presents:\\
 {\bf i.} In Sect.~\ref{SCFT} a short overview of the 
{\it spin-charge-family} theory is made. For more detailed discussions on the properties of the 
theory the reader is kindly invited to read Refs.~\cite{JMP2015,norma2014MatterAntimatter}, 
and in these references cited papers.  In App.~\ref{technique} the technique to represent 
spinors, used in this talk to explain the properties of the families and the family members, is
explained.\\ 
%In Apps.~(\ref{JMP2015IA}, \ref{IV}).
{\bf ii.} In Sect.~\ref{CPNandpropScm} the properties of the scalar fields, contributing to 
the electroweak break, are discussed, showing that all the scalar fields - the three singlets carrying 
the family members quantum numbers (interacting with both groups of four families) and the 
twice two triplets carrying the family quantum numbers (each pair of two triplets interacts with 
its own group of four families) - carry with respect to the scalar index the quantum numbers of the
weak and the hyper charges ($\pm \frac{1}{2}$, $\mp \frac{1}{2}$) as does the {\it standard 
model} higgs.\\
%27.06. 19:00
{\bf iii.} In Sect.~\ref{fourthfamily} couplings of  quarks to the scalar fields in the
 {\it spin-charge-family} theory are discussed in order to explain that the appearance of the fourth 
family might not contradict the observations. All the scalars with the space index $s=(7,8)$, 
Eq.~(\ref{eigentau1tau2}, \ref{faction}) carry the weak and the hyper charge of the {\it standard
model} higgs. The fourth family quarks couple stronger to the scalar fields which carry the family
quantum numbers, while the three observed families couple stronger to the scalar fields carrying 
the family members quantum numbers what causes a large differences in masses of the first
three families. 

Couplings to the scalar fields carrying the family quantum numbers
($\vec{\tilde{\tau}}^{1}$, $\vec{\tilde{N}}_{L}$), Eq.~(\ref{commonAi}) of  $u_{4}$ appear to 
contribute to these scalars with the opposite sign  than that of $d_{4}$, Fig.~\ref{Fig.1}, while the 
couplings of quarks to the scalars carrying the family members quantum numbers, $(Q,Q',Y')$, 
are determined besides by the coupling constants of the three singlets also by the eigenvalues of 
$ Q,Q',Y')$ of the family members $u_{i}$ and $d_{i}$. 

Correspondingly the contribution of the $u_{4}$ and $d_4$ quarks to the production of any 
superposition of the scalar fields carrying the family quantum numbers in  the quark-gluon fusion
is weakened, if masses of the scalars are high and these two family members 
have comparable masses.  The fourth family members 
$u_{4}$ and $d_{4}$ are expected to have comparable masses~\cite{gn2013,gn2015} if 
their masses are determined mostly by the  scalar fields which are the superposition of the scalar 
fields with the family quantum numbers~\cite{norma2016higgs}.\\ 
In this case mostly the top ($u_{3}$) quarks, the couplings of which to the superposition of the 
scalar fields with the family members quantum numbers are strong, contribute to the quark-gluon 
fusion.

% The scalars with the family members quantum numbers are responsible for the large 
%differences in masses of the lower three families of quarks, while their contribution to the masses 
%of the $u_{4}$ and $d_{4}$ quarks is much smaller. 

In the case of the weak (but not zero) contribution of the scalar fields 
with the family members quantum numbers to the $u_{4}$ and $d_{4}$ mass matrices, a mass 
eigenstate of the scalar fields which is mostly superposition of scalars with the family quantum 
numbers can be produced in the quark-gluon fusion. Since such a scalar 
field mostly couples to the fourth family quarks, and since their matrix elements to the 
lower three families are very small, the two photons  event, observed at the 
LHC~\cite{twophotonsATLAS, twophotonsCMS} at $\approx 750GeV$, might be the first observed 
signal that there are additional scalars, in agreement with the prediction of the 
{\it spin-charge-family} theory. The strong indirect signal that there are several scalar fields 
are the existence of the families and the Yukawa couplings of the higgs to the family members 
of the families.
%** Explain that gluon fusion depends on the energy and read the introduction , The fourth..
% and Conclusion again

%
\section{The spin-charge family theory, the starting action and assumptions}
\label{SCFT}

I present in this section the {\it assumptions} of the {\it spin-charge-family} theory, on which
the theory is built, following a lot the similar one from 
Refs.~\cite{norma2014MatterAntimatter,JMP2015}:

%\vspace{2mm}

\noindent
{\bf A i.} $\;\;$
In the action~\cite{norma2014MatterAntimatter,JMP,NBled2013,JMP2015} fermions 
$\psi$ carry in $d=(13+1)$ as the {\it internal degrees of freedom only two kinds of spins}  (no 
charges), which are determined by the two kinds of the Clifford algebra objects (there exist no 
additional Clifford algebra objects (\ref{gammatildegamma})) -
$\gamma^a$ and $\tilde{\gamma}^a$ -  and {\it interact correspondingly with the two kinds of
the spin connection fields} - $\omega_{ab \alpha}$ and $\tilde{\omega}_{ab \alpha}$, the 
{\it gauge fields of} $S^{ab}=\frac{i}{4}\,$ $(\gamma^a \gamma^b -\gamma^b 
\gamma^a)$ (the generators  of $SO(13,1)$) %(App.~\ref{generatorsandfields}) 
and  $\tilde{S}^{ab}=\frac{i}{4}\,$ $(\tilde{\gamma}^a \tilde{\gamma}^b - 
\tilde{\gamma}^b \tilde{\gamma}^a)$  (the generators of $\widetilde{SO}(13,1)$) 
%(App.~\ref{generatorsandfields}) 
-{\it and the} {\it vielbeins} $f^{\alpha}{}_{a}$.
\begin{eqnarray}
{\cal A}            \,  &=& \int \; d^dx \; E\;{\mathcal L}_{f} +    
               \int \; d^dx \; E\; (\alpha \,R + \tilde{\alpha} \, \tilde{R})\,,\nonumber\\
               %\end{eqnarray}
%
%\begin{eqnarray}
{\mathcal L}_f &=& \frac{1}{2}\, (\bar{\psi} \, \gamma^a p_{0a} \psi) + h.c., 
\nonumber\\
p_{0a }        &=& f^{\alpha}{}_a p_{0\alpha} + \frac{1}{2E}\, \{ p_{\alpha}, 
E f^{\alpha}{}_a\}_-,  \quad      p_{0\alpha} =  p_{\alpha}  - 
                    \frac{1}{2}  S^{ab} \omega_{ab \alpha} - 
                    \frac{1}{2}  \tilde{S}^{ab}   \tilde{\omega}_{ab \alpha},                   
\nonumber\\ 
R              &=&  \frac{1}{2} \, \{ f^{\alpha [ a} f^{\beta b ]} \;(\omega_{a b \alpha, \beta} 
- \omega_{c a \alpha}\,\omega^{c}{}_{b \beta}) \} + h.c. \;, 
\nonumber\\
\tilde{R}      &=&  \frac{1}{2} \, \{ f^{\alpha [ a} f^{\beta b ]} \;
(\tilde{\omega}_{a b \alpha,\beta} - 
\tilde{\omega}_{c a \alpha} \,\tilde{\omega}^{c}{}_{b \beta})\} + h.c.\;. 
\label{action}
\end{eqnarray}
Here~\footnote{$f^{\alpha}{}_{a}$ are inverted vielbeins to 
$e^{a}{}_{\alpha}$ with the properties $e^a{}_{\alpha} f^{\alpha}{\!}_b = \delta^a{\!}_b,\; 
e^a{\!}_{\alpha} f^{\beta}{\!}_a = \delta^{\beta}_{\alpha} $, $ E = \det(e^a{\!}_{\alpha}) $. 
Latin indices  
$a,b,..,m,n,..,s,t,..$ denote a tangent space (a flat index),
while Greek indices $\alpha, \beta,..,\mu, \nu,.. \sigma,\tau, ..$ denote an Einstein 
index (a curved index). Letters  from the beginning of both the alphabets
indicate a general index ($a,b,c,..$   and $\alpha, \beta, \gamma,.. $ ), 
from the middle of both the alphabets   
the observed dimensions $0,1,2,3$ ($m,n,..$ and $\mu,\nu,..$), indices from 
the bottom of the alphabets
indicate the compactified dimensions ($s,t,..$ and $\sigma,\tau,..$). 
We assume the signature $\eta^{ab} =
diag\{1,-1,-1,\cdots,-1\}$.} 
$f^{\alpha [a} f^{\beta b]}= f^{\alpha a} f^{\beta b} - f^{\alpha b} f^{\beta a}$.
%$S^{ab}$ and $\tilde{S}^{ab}$ are generators  of the 
%groups $SO(13,1)$ and $\widetilde{SO}(13,1)$, 
%respectively, %expressible by $\gamma^a$ and $\tilde{\gamma}^a$, 
$R$ and $\tilde{R}$ are the two scalars ($R$ is a curvature).
%, as it is presented in Sect.~\ref{lorentz} and in App.\ref{auxilliaryappendix}. 

%\vspace{2mm}

\noindent
{\bf A ii.} $\;\;$ 
The manifold $M^{(13+1)}$ breaks first into $M^{(7+1)}$ times $M^{(6)}$ (manifesting 
as $SO(7,1)$ $\times SU(3)$ $\times U(1)$), affecting both internal degrees of freedom - the 
one represented by (a superposition of) $S^{ab}$ and the one represented by (a superposition of) 
$\tilde{S}^{ab}$. Since the 
left handed (with respect to $M^{(7+1)}$) spinors couple differently to scalar (with respect to 
$M^{(7+1)}$) fields than the right handed ones, the break can leave massless and mass
protected $2^{((7+1)/2-1)}$ massless families (which decouple into twice four families). The 
rest of families get heavy
masses~\footnote{A toy model~\cite{DHN,DN012} was studied in $d=(5+1)$ with the same 
action as in Eq.~(\ref{action}). The break from $d=(5+1)$ to $d=(3+1) \times$ an almost
$S^{2}$ was studied. For a particular choice of vielbeins and for a class of spin connection fields 
the manifold $M^{(5+1)}$ breaks into $M^{(3+1)}$ times an almost $S^2$, while 
$2^{((3+1)/2-1)}$ families remain massless and mass protected. Equivalent assumption, 
its proof  is in progress, is made in the $d=(13+1)$ case.}.  

\noindent
{\bf A iii.} $\;\;$ 
The manifold $M^{(7+1)}$ breaks further into 
$M^{(3+1)} \times$  $M^{(4)}$.

\noindent
{\bf A iv.} $\;$ 
The scalar condensate (Table~\ref{Table con.}) of two right handed neutrinos with the family 
quantum numbers of one of the two groups of four families, brings masses of the scale of the
unification ($> 10^{16}$ GeV) to all the vector and scalar gauge fields, which 
interact with the condensate~\cite{norma2014MatterAntimatter}.  
 \begin{table}
 %\begin{center}
 %\begin{small}
\tbl{This table is taken from~\cite{norma2014MatterAntimatter}.
The condensate of the two right handed neutrinos $\nu_{R}$,  with the $VIII^{th}$ 
family quantum numbers (it might be as well that all the neutrinos with the family quantum 
numbers from (V-VIII), that is from the upper four families, contribute to the condensate), 
coupled to spin zero ($S^{03}=0=S^{12}$) and belonging to a triplet with respect to the 
generators $\tau^{2i}$, while $\tau^{13}=0$, is presented, together with its two partners. 
The right handed neutrino has $Q=0=Y$. The triplet carries $\tau^4=-1$,  $\tilde{\tau}^{13}=0$
$\tilde{\tau}^{23}=1$, $\tilde{\tau}^{4} =-1$, $\tilde{N}_{R}^3 = 1$, $\tilde{N}_{L}^3 = 0$,
$\tilde{Y}=0 $, $\tilde{Q}=0$. 
The family quantum numbers are presented in Table~\ref{Table III.}. 
%Table is taken from Ref.~\cite{norma2014MatterAntimatter}.
}
{ \begin{tabular}{c| r r r r | r r c c c }
 \hline
 state & $\tau^{23}$ &$\tau^{4}$ & $Y$&$Q$&
 $\tilde{\tau}^{23}$   &$\tilde{\tau}^{4}$&$\tilde{Y} $ & $\tilde{Q}$ & $\tilde{N}_{R}^{3}$
 \\
 \hline
 ${\bf (|\nu_{1 R}^{VIII}>_{1}\,|\nu_{2 R}^{VIII}>_{2})}$
 & $1$ & $-1$ & $0$ & $0$ & $1$&$-1$& $0$& $0$& $1$\\ 
 \hline
 $ (|\nu_{1 R}^{VIII}>_{1}|e_{2 R}^{VIII}>_{2})$
 & $0$ & $-1$ & $-1$& $-1$ & $1$&$-1$& $0$& $0$& $1$\\ 
 $ (|e_{1 R}^{VIII}>_{1}|e_{2 R}^{VIII}>_{2})$
 & $-1$& $-1$ & $-2$& $-2$ & $1$&$-1$& $0$& $0$& $1$\\ 
 \hline 
 \end{tabular}
}
 %\end{small}
 %\end{center}

\label{Table con.}
 \end{table} 

%\vspace{2mm}

\noindent
{\bf A v.} $\;\;$ 
There are nonzero vacuum expectation values of the scalar fields with the space index $s$ $=(7,8)$,  
conserving the electromagnetic and colour charge, which cause the electroweak break and bring
masses to all the fermions and to the heavy bosons.
%%%%%%%%%%%%%%%%%%%%%

\vspace{2mm}

{\it Comments on the assumptions}:

\vspace{2mm}

\noindent
{\bf C i.}  $\;\;$ 
The starting action contains all degrees of freedom, either for fermions or for bosons needed to 
manifest at low energy regime in $d=(3+1)$ all the vector and scalar gauge fields and  
the one family members as well as families of quarks and leptons as assumed by the 
{\it standard model}: 
 $\;$ {\bf a.}  $\;$ 
One representation of $SO(13,1)$ contains, if analyzed  with respect to the {\it standard model}
groups ($SO(3,1)\times SU(2) \times U(1)$ $ \times SU(3)$) all the members of one family 
 (Tables~\ref{Table so13+1.}--\ref{Table so13+1.b}), left and right handed, quarks and leptons (the right  handed
neutrino is one of the family members), with the quantum numbers required by the 
{\it standard model}~\footnote{It contains the left handed weak ($SU(2)_{I}$) 
charged and  $SU(2)_{II}$ chargeless colour triplet quarks and colourless leptons (neutrinos and 
electrons), and the right handed weak chargeless and $SU(2)_{II}$ charged coloured quarks and 
colourless leptons, as well as the right handed weak charged and $SU(2)_{II}$ chargeless colour 
anti-triplet anti-quarks and (anti)colourless anti-leptons, and the left handed weak chargeless and 
$SU(2)_{II}$ charged anti-quarks and anti-leptons. The anti-fermion states are reachable from 
the fermion states by the application of the discrete symmetry operator 
${\cal C}_{{\cal N}}$ ${\cal P}_{{\cal N}}$, presented in Ref.~\cite{discretesym}.}.
 $\;$ {\bf b.}  $\;$ The action explains the appearance of families due to the two kinds of 
the infinitesimal generators of groups: $S^{ab}$ and 
$\tilde{S}^{ab}$~\footnote{There are before the 
electroweak break two  decoupled groups of four massless families of quarks and leptons, in the 
fundamental representations of $\widetilde{SU}(2)_{R,\widetilde{SO}(3,1)}\times$ 
$ \widetilde{SU}(2)_{II,\widetilde{SO}(4)}$ and 
$\widetilde{SU}(2)_{L,\widetilde{SO}(3,1)}\times$ $ \widetilde{SU}(2)_{I,\widetilde{SO}(4)}$ 
groups, respectively - the subgroups of $\widetilde{SO}(3,1) $ and $\widetilde{SO}(4)$ 
(Table~\ref{Table III.}).
These eight families remain massless up to the electroweak break due to the "mass protection 
mechanism", that is due to the fact that the right handed members have no left handed 
partners with the same charges.}.
 $\;$ {\bf c.}  $\;$ The action explains the appearance of the gauge fields of the 
{\it standard model}~\cite{norma2014MatterAntimatter,JMP2015}.
(In Ref~\cite{JMP2015}, Sect. II. the proof is presented, that gauge 
fields can  in the Kaluza-Klein theories be equivalently represented by either the vielbeins or 
spin connection fields.)~\footnote{Before the electroweak break are all 
observable gauge fields massless: the gravity, the colour octet vector gauge fields (of the group 
$SU(3)$ from $SO(6)$), 
%Eq.~(\ref{so64})), 
the weak triplet vector gauge field  (of the group $SU(2)_{I}$ from $SO(4)$),
%Eq.~(\ref{so42})) 
and  the hyper singlet vector gauge field (a superposition of $U(1)$ from 
$SO(6)$ and the third component of $SU(2)_{II}$ triplet). % Eq.~(\ref{Aomegas})).
All are the superposition of the $f^{\alpha}{}_{c}$ $\omega_{ab \alpha}$ spinor gauge fields}.
$\;$ {\bf d.}  $\;$ It explains the appearance of the scalar higgs and Yukawa
 couplings~\footnote{There are scalar fields with the space index ($7,8$) 
and with respect to the space index with the weak and the hyper charge of the Higgs's scalar. 
%(Eq.~(\ref{checktau13Y})). 
They belong with respect to additional quantum numbers either to one of the two groups of 
two triplets, 
%Eqs.~(\ref{so1+3tilde}, \ref{so42tilde})
(either to one of the two triplets of the groups
$\widetilde{SU}(2)_{R\,\widetilde{SO}(3,1)}$ and $ \widetilde{SU}(2)_{II\,\widetilde{SO}(4)}$, 
or to one of the two triplets of the groups $\widetilde{SU}(2)_{L\,\widetilde{SO}(3,1)}$ and
 $ \widetilde{SU}(2)_{I\,\widetilde{SO}(4)}$, respectively), which couple through the family 
quantum numbers to one (the first two triplets) or to another (the second two triplets)  
group of four families - all are the superposition of $f^{\sigma}{}_{s} \,$ 
$\tilde{\omega}_{ab\sigma}$,
% (Eq.~(\ref{Atildeomegas})), 
or they belong to three singlets, 
the scalar gauge fields of ($Q,Q',Y'$), 
%(Eq.~(\ref{YQY'Q'andtilde})),
 which couple to the family 
members of both groups of families - they are the superposition of $f^{\sigma}{}_{s} \,$ 
$\omega_{ab\sigma}$.
 % (Eq.~(\ref{Aomegas})).
%
Both kinds of scalar fields determine the fermion masses (Eq.~(\ref{M0})), offering the 
explanation for the higgs, the Yukawa couplings and the heavy bosons masses.}.
% (Eq.~(\ref{bosonmassdoubletItriplets})). 
%
 $\;$ {\bf e.}  $\;$ The starting action contains also the additional $SU(2)_{II}$ (from $SO(4)$) 
%Eq.~(\ref{so42} 
vector gauge triplet (one of the components contributes to the hyper charge gauge fields as 
explained above), as well as the scalar fields  with the space index $s\in (5,6)$ and 
$t\in (9,10,\dots,14)$. All these fields gain masses of the scale of the condensate
(Table~\ref{Table con.}), which they interact with. They all are expressible with the superposition 
of $f^{\mu}{}_{m}\,\omega_{ab\mu}$ or 
of $f^{\mu}{}_{m}\,\tilde{\omega}_{ab\mu}$~\footnote{ 
In the case of free fields (if no spinor source, carrying their quantum numbers, is present) both
 $f^{\mu}{}_{m}\, \omega_{ab\mu}$ and $f^{\mu}{}_{m}\, \tilde{\omega}_{ab\mu}$ are 
expressible with vielbeins, 
%(Subsect.~\ref{spinconandvielbein}, Eqs.~(\ref{falphaa}, \ref{omegatildeabe})),
correspondingly only one kind of the three gauge fields are the 
propagating fields.}.

\noindent
{\bf C ii., C iii.} $\;\;$ There are many ways of breaking symmetries from $d=(13+1)$ to
$d=(3+1)$. The assumed breaks explain the connection between the weak and the hyper charge 
and the handedness of spinors, manifesting correspondingly the observed properties of the 
family members - the quarks and the leptons, left and right handed (Tables~\ref{Table so13+1.}--%
\ref{Table so13+1.b}~%
\footnote{The phases on Tables~\ref{Table so13+1.}--\ref{Table so13+1.b}  are all chosen to be 
$1$. In this talk 
 I multiplied some of the states by $(-1)$:  a. the states with the spin up 
($u_R$,   ${\bar u}_{R}$, ${\bar u}_{L}$) go to 
(-$u_R$, -$\bar{u}_{R}$,- $\bar{u}_{L}$), b. the states with the spin down 
($u_{L}$, $u_{R}$, $d_{L}$) go to 
(-$u_{L}$, -$u_{R}$, -$d_{L}$),  
correspondingly all the states have usual properties under the change of spin and under
 ${\cal C}_{{\cal N}}$ ${\cal P}_{{\cal N}}$.}.)  - and of the observed vector gauge fields.\\
After the break from $SO(13,1)$ to $SO(3,1)$ $\times SU(2) \times U(1) \times$ $SU(3)$ the 
anti-particles are accessible from particles by the application of the operator 
$\mathbb{C}_{{\cal N}}$ $\cdot {\cal P}_{{\cal N}}$, as explained~%
\footnote{The discrete symmetry operator $\mathbb{C}_{{\cal N}}$ $\cdot {\cal P}_{{\cal N}}$, 
Refs.~\cite{discretesym,TDN}, does not contain $\tilde{\gamma}^a$'s degrees of freedom. 
To each family member there corresponds the anti-member, with the same family quantum 
number.} in Refs.~\cite{discretesym}.

\noindent
{\bf C iv.} $\;\;$ 
It is the condensate (Table~\ref{Table con.}) of two right handed neutrinos with the quantum 
numbers of one group of four families, which makes massive all the scalar gauge fields (with the 
index ($5,6,7,8$), as well as those with the index $(9,\dots,14)$) and the vector gauge fields, 
manifesting nonzero $\tau^{4}$, $\tau^{23}$, %$Q$ ,$Y$, 
$\tilde{\tau}^{4}$, $\tilde{\tau}^{23}$, %$\tilde{Q}$ ,$\tilde{Y}$, 
$\tilde{N}^{3}_{R}$~% 
\cite{norma2014MatterAntimatter,JMP2015}. Only the vector gauge fields of 
$Y$ $(U(1))$, $\vec{\tau}^{3}$ $(SU(3))$ and $\vec{\tau}^1$ $(SU(2))$ remain massless, 
since they do not interact with the condensate.

\noindent
{\bf C v.} $\;\;$ 
At the electroweak break the scalar fields with the space index $s=(7,8)$ 
%(conserving the  electromagnetic and colour charges)
 - originating in 
$\tilde{\omega}_{abs}$, % (Eq.~(\ref{Atildeomegas}) 
as well as some superposition of
$\omega_{s' s" s}$ with the quantum numbers ($Q,Q',Y'$), %Eq.~(\ref{Aomegas}), 
conserving the colour and the electromagnetic charge - change their mutual interaction, and 
gaining nonzero 
vacuum expectation values change correspondingly also their masses. They contribute to mass 
matrices of twice the four families, as well as to the masses of the heavy vector bosons. 

All the rest scalar fields keep masses of the scale of the condensate and are correspondingly  
unobservable in the low energy regime.

The fourth family to the observed three ones is {\it  predicted to be observed} at the LHC. Its 
properties are under consideration~\cite{gn2013,gn2015}, the {\it baryons of the stable family of 
the upper four families offer the explanation for the dark matter}~\cite{gn}. The {\it 
triplet and anti-triplet scalar fields}  contribute together with the condensate to the
 {\it matter/anti-matter asymmetry}.

\vspace{2mm}

Let us (formally)  rewrite that part of the action of Eq.(\ref{action}), which determines the 
spinor degrees of freedom, in the way that we can clearly see that the action does  in the low 
energy regime manifest by the {\it standard model} required 
degrees of freedom of the fermions, vector and scalar gauge fields~\cite{NBled2013,NBled2012,%
JMP,portoroz03,JMP2015,pikanorma,portoroz03,norma92,norma93,norma94,norma95,gmdn07,gn,gn2013}.
\begin{eqnarray}
\label{faction}
{\mathcal L}_f &=&  \bar{\psi}\gamma^{m} (p_{m}- \sum_{A,i}\; g^{A}\tau^{Ai} 
A^{Ai}_{m}) \psi + \nonumber\\
               & &  \{ \sum_{s=7,8}\;  \bar{\psi} \gamma^{s} p_{0s} \; \psi \} + \nonumber\\ 
& & \{ \sum_{t=5,6,9,\dots, 14}\;  \bar{\psi} \gamma^{t} p_{0t} \; \psi \}
%+\nonumber\\       & &       {\rm the \;rest}
\,, 
\end{eqnarray}
where $p_{0s} =  p_{s}  - \frac{1}{2}  S^{s' s"} \omega_{s' s" s} - 
                    \frac{1}{2}  \tilde{S}^{ab}   \tilde{\omega}_{ab s}$, 
$p_{0t}   =    p_{t}  - \frac{1}{2}  S^{t' t"} \omega_{t' t" t} - 
                    \frac{1}{2}  \tilde{S}^{ab}   \tilde{\omega}_{ab t}$,                    
with $ m \in (0,1,2,3)$, $s \in (7,8),\, (s',s") \in (5,6,7,8)$, $(a,b)$ (appearing in
 $\tilde{S}^{ab}$) run within  either $ (0,1,2,3)$ or $ (5,6,7,8)$, $t$ runs $ \in (5,\dots,14)$, 
$(t',t")$ run either $ \in  (5,6,7,8)$ or $\in (9,10,\dots,14)$. 
The spinor function $\psi$ represents all family members of all the $2^{\frac{7+1}{2}-1}=8$ 
families.

The first line of Eq.~(\ref{faction}) determines (in $d=(3+1)$) the kinematics and dynamics of
spinor (fermion) fields, coupled to the vector gauge fields. The generators $\tau^{Ai} $ of the charge 
groups are expressible  
%in Eq.~(\ref{taucom}), 
in terms of $S^{ab}$ through the complex coefficients $c^{Ai}{ }_{ab}$~\footnote{ 
%$\vec{N}_{\pm}(= \vec{N}_{(L,R)}): = \,\frac{1}{2} (S^{23}\pm i S^{01},S^{31}\pm i S^{02}, 
%S^{12}\pm i S^{03}$, 
$\vec{\tau}^{1}:=\frac{1}{2} (S^{58}-  S^{67}, \,S^{57} + S^{68}, \,S^{56}-  S^{78} )\,,
  \vec{\tau}^{2}:=\frac{1}{2} (S^{58}+  S^{67}, \,S^{57} - S^{68}, \,S^{56}+  S^{78} )$,
$ \vec{\tau}^{3}: = \frac{1}{2} \,\{  S^{9\;12} - S^{10\;11} \,,
  S^{9\;11} + S^{10\;12} ,\, S^{9\;10} - S^{11\;12}\,, S^{9\;14} -  S^{10\;13} ,\,      
  S^{9\;13} + S^{10\;14} \,,  S^{11\;14} -  S^{12\;13}\,,  S^{11\;13} +  S^{12\;14},\, 
 \frac{1}{\sqrt{3}} ( S^{9\;10} + S^{11\;12} -  2 S^{13\;14})\}\,$,
 $\tau^{4}: = -\frac{1}{3}(S^{9\;10} + S^{11\;12} + S^{13\;14})$.
After the electroweak break the charges 
$ Y:= \tau^{4} + \tau^{23}\,, Y':= -\tau^{4} \tan^2\vartheta_2 + \tau^{23}\,,
  Q: =  \tau^{13} + Y\,, Q':= -Y \tan^2\vartheta_1 + \tau^{13}$ manifest.  $\theta_1$ is 
the electroweak angle,  breaking $SU(2)_{I}$, $\theta_2$ is the angle of the break of the 
$SU(2)_{II}$ from $SU(2)_{I}\times SU(2)_{II}$.},
 % \tilde{Y}:= \tilde{\tau}^{4} + \tilde{\tau}^{23}\,, \tilde{Y'}:= -\tilde{\tau}^{4} 
  %\tan^2 \vartheta_2 + \tilde{\tau}^{23}\,,
  %\tilde{Q}:= \tilde{Y} + \tilde{\tau}^{13}\,, \tilde{Q'}= -\tilde{Y} \tan^2 \vartheta_1 
 % + \tilde{\tau}^{13}$
%Eqs.~(\ref{so42}, \ref{so64}, 
%\ref{YQY'Q'andtilde} 
%
\begin{eqnarray}
\tau^{Ai} = \sum_{a,b} \;c^{Ai}{ }_{ab} \; S^{ab}\,,
%\nonumber\\ 
%\{\tau^{Ai}, \tau^{Bj}\}_- = i \delta^{AB} f^{Aijk} \tau^{Ak}\,.
\label{tau}
\end{eqnarray}
fulfilling the commutation relations 
\begin{eqnarray}
%\tau^{Ai} = \sum_{a,b} \;c^{Ai}{ }_{ab} \; S^{ab}\,,
%\nonumber\\ 
\{\tau^{Ai}, \tau^{Bj}\}_- = i \delta^{AB} f^{Aijk} \tau^{Ak}\,.
\label{taucom}
\end{eqnarray}
They represent the colour ($\tau^{3i}$), the weak $(\tau^{1i})$ and the hyper ($Y$) 
charges (as well as the $SU(2)_{II}$ ($\tau^{2i}$) and
$\tau^{4}$ charges, the gauge fields of which gain masses interacting with  the condensate, 
Table~\ref{Table con.},  leaving massless only the hyper charge vector gauge field). The 
corresponding vector gauge fields  $A^{Ai}_{m}$ are expressible with the spin connection fields
 $\omega_{stm}$, 
% (Eqs.~(\ref{Aomegasvector}, \ref{Atildeomegasvector})), 
with $(s,t)$  either in $ (5,6,7,8)$ or in
$ (9,\dots,14)$, in agreement with the assumptions {\bf A ii.} and {\bf A iii.}. I demonstrate in 
Ref.~\cite{JMP2015} the equivalence between  the usual 
Kaluza-Klein procedure leading to the vector gauge fields  through the vielbeins and the procedure 
with the spin connections proposed by the {\it spin-charge-family} theory. 

All vector gauge fields, appearing in the first line of Eq.~(\ref{faction}), except 
$A^{2 \pm}_{m}$ and $A^{Y'}_{m}$ ($=\cos \vartheta_{2} \,A^{23}_{m} - 
\sin \vartheta_{2}\, A^{4}_{m}$, $Y'$  and $\tau^{4}$ are defined 
in~\footnote{$Y':= -\tau^{4} \tan^2\vartheta_2 + \tau^{23}$, $\tau^{4} = -\frac{1}{3}(%
S^{9\,10}+ S^{11\,12} + S^{13\,14})$.}),  
%Eq.~(\ref{YQY'Q'andtilde}),  
%in Eq.~(\ref{so64})),
are massless before the electroweak break. 
$\vec{A}^{3}_{m}$ carries the colour charge $SU(3)$ (originating in $SO(6)$),
$\vec{A}^{1}_{m}$ carries the weak charge $SU(2)_{I}$ ($SU(2)_{I}$ and  $SU(2)_{II}$ 
are the subgroups of $SO(4)$) and $A^{Y}_{m}$ ($= \sin \vartheta_{2}\, A^{23}_{m} +
\cos \vartheta_{2} \,A^{4}_{m}\,$) carries %$Y$ is defined in Eq.~(\ref{YQY'Q'andtilde}),
the corresponding $U(1)$ charge ($Y=\tau^{23} + \tau^{4}$, $\tau^{4}$ originates in $SO(6)$  
%$A^{4}_{m}$ is defined in Eq.~(\ref{Aomegas}) if the scalar space index $s$ is replaced by 
%the space vector index $m$,   
and $\tau^{23}$ is the third component of the second $SU(2)_{II}$ group, $A^{4}_{m}$ and 
$\vec{A}^{2}_{m}$ are the corresponding vector gauge fields). 
The fields $A^{2 \pm}_{m}$ and $A^{Y'}_{m}$ get masses of the 
order of the condensate scale through the interaction with the condensate of the two right
handed neutrinos with the quantum numbers of one of the group of four families 
(the assumption {\bf A iv.}, Table~\ref{Table con.}). (See Ref.~\cite{JMP2015}.)

%\begin{tiny} 
\begin{sidewaystable}
%\topcaption{%\label{Table so13+1.}%
%\tiny{
\tbl{The left handed ($\Gamma^{(13,1)} = -1$)  ($ = \Gamma^{(7,1)} \times \Gamma^{(6)}$) 
multiplet of spinors - the members of the $SO(13,1)$ group, 
manifesting the subgroup $SO(7,1)$ - of the colour charged quarks and anti-quarks and the colourless 
leptons and anti-leptons, is presented in the massless basis using the technique presented in
App.~\ref{technique}. %~\cite{snmb:hn02hn03}. 
It contains the left handed  ($\Gamma^{(3,1)}=-1$) weak charged  ($\tau^{13}=\pm \frac{1}{2}$) 
%,$\vec{\tau}^{1}= \frac{1}{2}(S^{58}- S^{67},S^{57}+ S^{68},S^{56}- S^{78} )$) 
and $SU(2)_{II}$ chargeless ($\tau^{23}=0$) 
%, $\vec{\tau}^{2}= \frac{1}{2}(S^{58}+ S^{67},S^{57}- S^{68},S^{56}+ S^{78})$) 
quarks and the right handed weak chargeless and $SU(2)_{II}$ charged ($\tau^{23}=\pm \frac{1}{2}$) 
quarks of three colours  ($c^i$ $= (\tau^{33}, \tau^{38})$) %$((\frac{1}{2},\frac{1}{2\sqrt{3}}, 
%(-\frac{1}{2},\frac{1}{2\sqrt{3}}, (0,-\frac{1}{\sqrt{3}}) $), 
%$\vec{\tau}^{3}= \frac{1}{2}(S^{9\,12}- S^{10\,11},S^{9\,11}+ S^{10\,12},S^{9\,10}- S^{11\,12},$
%$S^{9\,14}- S^{10\,13},S^{9\,13}+ S^{10\,14},S^{11\,14}- S^{12\,13},$
%$S^{11\,13}+ S^{12\,14},\frac{1}{\sqrt{3}}(S^{9\,10}+ S^{11\,12} - 2S^{13\,14})$), 
with the "spinor" charge ($\tau^{4}=\frac{1}{6}$) %$=-\frac{1}{3}(S^{9\,10}+ S^{11\,12}+ S^{13\,14})$)
and the colourless left handed weak charged leptons and the right handed weak chargeless leptons
with the "spinor" charge ($\tau^{4}=-\frac{1}{2}$).  
$ S^{12}$ defines the ordinary spin %(which can also be read directly from the basic vector, both
%vectors  with both spins, 
$\pm \frac{1}{2}$. It contains also the states of opposite charges, reachable from particle states 
by the application of the discrete symmetry operator 
${\cal C}_{{\cal N}}$ ${\cal P}_{{\cal N}}$, presented in Refs.~\cite{discretesym,TDN}.
%
%Additional notations are presented in Table~\ref{Table II.}.
%The representation contains, similarly as in the 
%case of the toy model~Table\ref{Table I.}, the triplet and the anti-triplet representations of quarks and 
%similarly colourless and anti-colourless representation of leptons. $\tau^{4}$ defines the $U(1)_{II}$ charge 
%The hypercharge ($Y= \tau^4 + \tau^{23}$) and the electromagnetic charge $Q= Y + \tau^{13}$ are also presented. 
The vacuum state, 
%$|vac>_{fam}$, 
on which the nilpotents and projectors operate, is not shown. 
%The basis is the massless one. 
The reader can find this  Weyl representation also in Refs.~\cite{norma2014MatterAntimatter,%
Portoroz03,JMP}. Table is separated into three parts.
%Two anti-octets of anti-quarks of the rest two anti-triplet colours follow from the presented one by substituting 
%the part $||\stackrel{9 \;10}{[-]}\;\;\stackrel{11\;12}{(+)}\;\;\stackrel{13\;14}{(+)}$ of the basis by 
%$||\stackrel{9 \;10}{(+)}\;\;\stackrel{11\;12}{[-]}\;\;\stackrel{13\;14}{(+)}$ for $\bar{q}_{L,R}^{\bar{c^2}}$
%and by $||\stackrel{9 \;10}{(+)}\;\;\stackrel{11\;12}{(+)}\;\;\stackrel{13\;14}{[-]}$ for 
%$\bar{q}_{L,R}^{\bar{c^3}}$. Correspondingly the charges are ($\frac{1}{2}, - \frac{1}{2 \sqrt{3}}$) and 
%($0,  \frac{1}{ \sqrt{3}}$), respectively.
%Left handed antiquarks and antileptons are weak chargeless and carry opposite charges.}%}
}
%\label{Table so13+1.}
%\end{tiny}
%\tablehead{
{\begin{tabular}{r c |c | c c|c c c|c c c |r r}
\hline
i&$$&$|^a\psi_i>$&$\Gamma^{(3,1)}$&$ S^{12}$&$\Gamma^{(4)}$&
$\tau^{13}$&$\tau^{23}$&$\tau^{33}$&$\tau^{38}$&$\tau^{4}$&$Y$&$Q$\\
\hline
&& ${\rm (Anti)octet},\,\Gamma^{(1,7)} = (-1)\,1\,, \,\Gamma^{(6)} = (1)\,-1$&&&&&&& \\
&& ${\rm of \;(anti) quarks \;and \;(anti)leptons}$&&&&&&&\\
\hline\hline 
%\tabletail{\hline \multicolumn{13}{r}{\emph{Continued on next page}}\\}
%\tablelasttail{\hline}
%\begin{table}
%\begin{small}
%\begin{tiny}
%\begin{center}
%\tiny{
%\begin{supertabular}{|r|c||c||c|c||c|c|c||c|c|c||r|r|}
1&$ u_{R}^{c1}$&$ \stackrel{03}{(+i)}\,\stackrel{12}{(+)}|
\stackrel{56}{(+)}\,\stackrel{78}{(+)}
||\stackrel{9 \;10}{(+)}\;\;\stackrel{11\;12}{(-)}\;\;\stackrel{13\;14}{(-)} $ &1&$\frac{1}{2}$&1&0&
$\frac{1}{2}$&$\frac{1}{2}$&$\frac{1}{2\,\sqrt{3}}$&$\frac{1}{6}$&$\frac{2}{3}$&$\frac{2}{3}$\\
\hline 
2&$u_{R}^{c1}$&$\stackrel{03}{[-i]}\,\stackrel{12}{[-]}|\stackrel{56}{(+)}\,\stackrel{78}{(+)}
||\stackrel{9 \;10}{(+)}\;\;\stackrel{11\;12}{(-)}\;\;\stackrel{13\;14}{(-)}$&1&$-\frac{1}{2}$&1&0&
$\frac{1}{2}$&$\frac{1}{2}$&$\frac{1}{2\,\sqrt{3}}$&$\frac{1}{6}$&$\frac{2}{3}$&$\frac{2}{3}$\\
\hline
3&$d_{R}^{c1}$&$\stackrel{03}{(+i)}\,\stackrel{12}{(+)}|\stackrel{56}{[-]}\,\stackrel{78}{[-]}
||\stackrel{9 \;10}{(+)}\;\;\stackrel{11\;12}{(-)}\;\;\stackrel{13\;14}{(-)}$&1&$\frac{1}{2}$&1&0&
$-\frac{1}{2}$&$\frac{1}{2}$&$\frac{1}{2\,\sqrt{3}}$&$\frac{1}{6}$&$-\frac{1}{3}$&$-\frac{1}{3}$\\
\hline 
4&$ d_{R}^{c1} $&$\stackrel{03}{[-i]}\,\stackrel{12}{[-]}|
\stackrel{56}{[-]}\,\stackrel{78}{[-]}
||\stackrel{9 \;10}{(+)}\;\;\stackrel{11\;12}{(-)}\;\;\stackrel{13\;14}{(-)} $&1&$-\frac{1}{2}$&1&0&
$-\frac{1}{2}$&$\frac{1}{2}$&$\frac{1}{2\,\sqrt{3}}$&$\frac{1}{6}$&$-\frac{1}{3}$&$-\frac{1}{3}$\\
\hline
5&$d_{L}^{c1}$&$\stackrel{03}{[-i]}\,\stackrel{12}{(+)}|\stackrel{56}{[-]}\,\stackrel{78}{(+)}
||\stackrel{9 \;10}{(+)}\;\;\stackrel{11\;12}{(-)}\;\;\stackrel{13\;14}{(-)}$&-1&$\frac{1}{2}$&-1&
$-\frac{1}{2}$&0&$\frac{1}{2}$&$\frac{1}{2\,\sqrt{3}}$&$\frac{1}{6}$&$\frac{1}{6}$&$-\frac{1}{3}$\\
\hline
6&$d_{L}^{c1} $&$\stackrel{03}{(+i)}\,\stackrel{12}{[-]}|\stackrel{56}{[-]}\,\stackrel{78}{(+)}
||\stackrel{9 \;10}{(+)}\;\;\stackrel{11\;12}{(-)}\;\;\stackrel{13\;14}{(-)} $&-1&$-\frac{1}{2}$&-1&
$-\frac{1}{2}$&0&$\frac{1}{2}$&$\frac{1}{2\,\sqrt{3}}$&$\frac{1}{6}$&$\frac{1}{6}$&$-\frac{1}{3}$\\
\hline
7&$ u_{L}^{c1}$&$\stackrel{03}{[-i]}\,\stackrel{12}{(+)}|\stackrel{56}{(+)}\,\stackrel{78}{[-]}
||\stackrel{9 \;10}{(+)}\;\;\stackrel{11\;12}{(-)}\;\;\stackrel{13\;14}{(-)}$ &-1&$\frac{1}{2}$&-1&
$\frac{1}{2}$&0 &$\frac{1}{2}$&$\frac{1}{2\,\sqrt{3}}$&$\frac{1}{6}$&$\frac{1}{6}$&$\frac{2}{3}$\\
\hline
8&$u_{L}^{c1}$&$\stackrel{03}{(+i)}\,\stackrel{12}{[-]}|\stackrel{56}{(+)}\,\stackrel{78}{[-]}
||\stackrel{9 \;10}{(+)}\;\;\stackrel{11\;12}{(-)}\;\;\stackrel{13\;14}{(-)}$&-1&$-\frac{1}{2}$&-1&
$\frac{1}{2}$&0&$\frac{1}{2}$&$\frac{1}{2\,\sqrt{3}}$&$\frac{1}{6}$&$\frac{1}{6}$&$\frac{2}{3}$\\
\hline\hline
%\shrinkheight{0.4\textheight}
9&$ u_{R}^{c2}$&$ \stackrel{03}{(+i)}\,\stackrel{12}{(+)}|
\stackrel{56}{(+)}\,\stackrel{78}{(+)}
||\stackrel{9 \;10}{[-]}\;\;\stackrel{11\;12}{[+]}\;\;\stackrel{13\;14}{(-)} $ &1&$\frac{1}{2}$&1&0&
$\frac{1}{2}$&$-\frac{1}{2}$&$\frac{1}{2\,\sqrt{3}}$&$\frac{1}{6}$&$\frac{2}{3}$&$\frac{2}{3}$\\
\hline 
10&$u_{R}^{c2}$&$\stackrel{03}{[-i]}\,\stackrel{12}{[-]}|\stackrel{56}{(+)}\,\stackrel{78}{(+)}
||\stackrel{9 \;10}{[-]}\;\;\stackrel{11\;12}{[+]}\;\;\stackrel{13\;14}{(-)}$&1&$-\frac{1}{2}$&1&0&
$\frac{1}{2}$&$-\frac{1}{2}$&$\frac{1}{2\,\sqrt{3}}$&$\frac{1}{6}$&$\frac{2}{3}$&$\frac{2}{3}$\\
\hline
$\cdots$&&&&&&&&&&&&\\
\hline % Correct also these parts as there are the above two candidates
\hline
\end{tabular}}
\label{Table so13+1.}
\end{sidewaystable}

\begin{sidewaystable}
\tbl{Continuation of Table\ref{Table so13+1.}}
{\begin{tabular}{r c |c | c c|c c c|c c c |r r}
%{\begin{tabular}{r c |c | c c|c c c|c c c |r r}
\hline
i&$$&$|^a\psi_i>$&$\Gamma^{(3,1)}$&$ S^{12}$&$\Gamma^{(4)}$&
$\tau^{13}$&$\tau^{23}$&$\tau^{33}$&$\tau^{38}$&$\tau^{4}$&$Y$&$Q$\\
\hline
&& ${\rm (Anti)octet},\,\Gamma^{(1,7)} = (-1)\,1\,, \,\Gamma^{(6)} = (1)\,-1$&&&&&&& \\
&& ${\rm of \;(anti) quarks \;and \;(anti)leptons}$&&&&&&&\\
\hline\hline 
\hline\hline
17&$ u_{R}^{c3}$&$ \stackrel{03}{(+i)}\,\stackrel{12}{(+)}|
\stackrel{56}{(+)}\,\stackrel{78}{(+)}
||\stackrel{9 \;10}{[-]}\;\;\stackrel{11\;12}{(-)}\;\;\stackrel{13\;14}{[+]} $ &1&$\frac{1}{2}$&1&0&
$\frac{1}{2}$&$0$&$-\frac{1}{\sqrt{3}}$&$\frac{1}{6}$&$\frac{2}{3}$&$\frac{2}{3}$\\
\hline 
18&$u_{R}^{c3}$&$\stackrel{03}{[-i]}\,\stackrel{12}{[-]}|\stackrel{56}{(+)}\,\stackrel{78}{(+)}
||\stackrel{9 \;10}{[-]}\;\;\stackrel{11\;12}{(-)}\;\;\stackrel{13\;14}{[+]}$&1&$-\frac{1}{2}$&1&0&
$\frac{1}{2}$&$0$&$-\frac{1}{\sqrt{3}}$&$\frac{1}{6}$&$\frac{2}{3}$&$\frac{2}{3}$\\
\hline
$\cdots$&&&&&&&&&&&&\\
\hline\hline
25&$ \nu_{R}$&$ \stackrel{03}{(+i)}\,\stackrel{12}{(+)}|
\stackrel{56}{(+)}\,\stackrel{78}{(+)}
||\stackrel{9 \;10}{(+)}\;\;\stackrel{11\;12}{[+]}\;\;\stackrel{13\;14}{[+]} $ &1&$\frac{1}{2}$&1&0&
$\frac{1}{2}$&$0$&$0$&$-\frac{1}{2}$&$0$&$0$\\
\hline 
26&$\nu_{R}$&$\stackrel{03}{[-i]}\,\stackrel{12}{[-]}|\stackrel{56}{(+)}\,\stackrel{78}{(+)}
||\stackrel{9 \;10}{(+)}\;\;\stackrel{11\;12}{[+]}\;\;\stackrel{13\;14}{[+]}$&1&$-\frac{1}{2}$&1&0&
$\frac{1}{2}$ &$0$&$0$&$-\frac{1}{2}$&$0$&$0$\\
\hline
27&$e_{R}$&$\stackrel{03}{(+i)}\,\stackrel{12}{(+)}|\stackrel{56}{[-]}\,\stackrel{78}{[-]}
||\stackrel{9 \;10}{(+)}\;\;\stackrel{11\;12}{[+]}\;\;\stackrel{13\;14}{[+]}$&1&$\frac{1}{2}$&1&0&
$-\frac{1}{2}$&$0$&$0$&$-\frac{1}{2}$&$-1$&$-1$\\
\hline 
28&$ e_{R} $&$\stackrel{03}{[-i]}\,\stackrel{12}{[-]}|
\stackrel{56}{[-]}\,\stackrel{78}{[-]}
||\stackrel{9 \;10}{(+)}\;\;\stackrel{11\;12}{[+]}\;\;\stackrel{13\;14}{[+]} $&1&$-\frac{1}{2}$&1&0&
$-\frac{1}{2}$&$0$&$0$&$-\frac{1}{2}$&$-1$&$-1$\\
\hline
29&$e_{L}$&$\stackrel{03}{[-i]}\,\stackrel{12}{(+)}|\stackrel{56}{[-]}\,\stackrel{78}{(+)}
||\stackrel{9 \;10}{(+)}\;\;\stackrel{11\;12}{[+]}\;\;\stackrel{13\;14}{[+]}$&-1&$\frac{1}{2}$&-1&
$-\frac{1}{2}$&0&$0$&$0$&$-\frac{1}{2}$&$-\frac{1}{2}$&$-1$\\
\hline
30&$e_{L} $&$\stackrel{03}{(+i)}\,\stackrel{12}{[-]}|\stackrel{56}{[-]}\,\stackrel{78}{(+)}
||\stackrel{9 \;10}{(+)}\;\;\stackrel{11\;12}{[+]}\;\;\stackrel{13\;14}{[+]} $&-1&$-\frac{1}{2}$&-1&
$-\frac{1}{2}$&0&$0$&$0$&$-\frac{1}{2}$&$-\frac{1}{2}$&$-1$\\
\hline
31&$ \nu_{L}$&$\stackrel{03}{[-i]}\,\stackrel{12}{(+)}|\stackrel{56}{(+)}\,\stackrel{78}{[-]}
||\stackrel{9 \;10}{(+)}\;\;\stackrel{11\;12}{[+]}\;\;\stackrel{13\;14}{[+]}$ &-1&$\frac{1}{2}$&-1&
$\frac{1}{2}$&0 &$0$&$0$&$-\frac{1}{2}$&$-\frac{1}{2}$&$0$\\
\hline
32&$\nu_{L}$&$\stackrel{03}{(+i)}\,\stackrel{12}{[-]}|\stackrel{56}{(+)}\,\stackrel{78}{[-]}
||\stackrel{9 \;10}{(+)}\;\;\stackrel{11\;12}{[+]}\;\;\stackrel{13\;14}{[+]}$&-1&$-\frac{1}{2}$&-1&
$\frac{1}{2}$&0&$0$&$0$&$-\frac{1}{2}$&$-\frac{1}{2}$&$0$\\
\hline\hline
\end{tabular}}
\label{Table so13+1.a}
\end{sidewaystable}

\begin{sidewaystable}
\tbl{Continuation of Table\ref{Table so13+1.}}
{\begin{tabular}{r c |c | c c|c c c|c c c |r r}
%{\begin{tabular}{|r|c||c||c|c||c|c|c||c|c|c||r|r|}
\hline
i&$$&$|^a\psi_i>$&$\Gamma^{(3,1)}$&$ S^{12}$&$\Gamma^{(4)}$&
$\tau^{13}$&$\tau^{23}$&$\tau^{33}$&$\tau^{38}$&$\tau^{4}$&$Y$&$Q$\\
\hline
&& ${\rm (Anti)octet},\,\Gamma^{(1,7)} = (-1)\,1\,, \,\Gamma^{(6)} = (1)\,-1$&&&&&&& \\
&& ${\rm of \;(anti) quarks \;and \;(anti)leptons}$&&&&&&&\\
\hline\hline 
33&$ \bar{d}_{L}^{\bar{c1}}$&$ \stackrel{03}{[-i]}\,\stackrel{12}{(+)}|
\stackrel{56}{(+)}\,\stackrel{78}{(+)}
||\stackrel{9 \;10}{[-]}\;\;\stackrel{11\;12}{[+]}\;\;\stackrel{13\;14}{[+]} $ &-1&$\frac{1}{2}$&1&0&
$\frac{1}{2}$&$-\frac{1}{2}$&$-\frac{1}{2\,\sqrt{3}}$&$-\frac{1}{6}$&$\frac{1}{3}$&$\frac{1}{3}$\\
\hline 
34&$\bar{d}_{L}^{\bar{c1}}$&$\stackrel{03}{(+i)}\,\stackrel{12}{[-]}|\stackrel{56}{(+)}\,\stackrel{78}{(+)}
||\stackrel{9 \;10}{[-]}\;\;\stackrel{11\;12}{[+]}\;\;\stackrel{13\;14}{[+]}$&-1&$-\frac{1}{2}$&1&0&
$\frac{1}{2}$&$-\frac{1}{2}$&$-\frac{1}{2\,\sqrt{3}}$&$-\frac{1}{6}$&$\frac{1}{3}$&$\frac{1}{3}$\\
\hline
35&$\bar{u}_{L}^{\bar{c1}}$&$\stackrel{03}{[-i]}\,\stackrel{12}{(+)}|\stackrel{56}{[-]}\,\stackrel{78}{[-]}
||\stackrel{9 \;10}{[-]}\;\;\stackrel{11\;12}{[+]}\;\;\stackrel{13\;14}{[+]}$&-1&$\frac{1}{2}$&1&0&
$-\frac{1}{2}$&$-\frac{1}{2}$&$-\frac{1}{2\,\sqrt{3}}$&$-\frac{1}{6}$&$-\frac{2}{3}$&$-\frac{2}{3}$\\
\hline
36&$ \bar{u}_{L}^{\bar{c1}} $&$\stackrel{03}{(+i)}\,\stackrel{12}{[-]}|
\stackrel{56}{[-]}\,\stackrel{78}{[-]}
||\stackrel{9 \;10}{[-]}\;\;\stackrel{11\;12}{[+]}\;\;\stackrel{13\;14}{[+]} $&-1&$-\frac{1}{2}$&1&0&
$-\frac{1}{2}$&$-\frac{1}{2}$&$-\frac{1}{2\,\sqrt{3}}$&$-\frac{1}{6}$&$-\frac{2}{3}$&$-\frac{2}{3}$\\
\hline
37&$\bar{d}_{R}^{\bar{c1}}$&$\stackrel{03}{(+i)}\,\stackrel{12}{(+)}|\stackrel{56}{(+)}\,\stackrel{78}{[-]}
||\stackrel{9 \;10}{[-]}\;\;\stackrel{11\;12}{[+]}\;\;\stackrel{13\;14}{[+]}$&1&$\frac{1}{2}$&-1&
$\frac{1}{2}$&0&$-\frac{1}{2}$&$-\frac{1}{2\,\sqrt{3}}$&$-\frac{1}{6}$&$-\frac{1}{6}$&$\frac{1}{3}$\\
\hline
38&$\bar{d}_{R}^{\bar{c1}} $&$\stackrel{03}{[-i]}\,\stackrel{12}{[-]}|\stackrel{56}{(+)}\,\stackrel{78}{[-]}
||\stackrel{9 \;10}{[-]}\;\;\stackrel{11\;12}{[+]}\;\;\stackrel{13\;14}{[+]} $&1&$-\frac{1}{2}$&-1&
$\frac{1}{2}$&0&$-\frac{1}{2}$&$-\frac{1}{2\,\sqrt{3}}$&$-\frac{1}{6}$&$-\frac{1}{6}$&$\frac{1}{3}$\\
\hline
39&$ \bar{u}_{R}^{\bar{c1}}$&$\stackrel{03}{(+i)}\,\stackrel{12}{(+)}|\stackrel{56}{[-]}\,\stackrel{78}{(+)}
||\stackrel{9 \;10}{[-]}\;\;\stackrel{11\;12}{[+]}\;\;\stackrel{13\;14}{[+]}$ &1&$\frac{1}{2}$&-1&
$-\frac{1}{2}$&0 &$-\frac{1}{2}$&$-\frac{1}{2\,\sqrt{3}}$&$-\frac{1}{6}$&$-\frac{1}{6}$&$-\frac{2}{3}$\\
\hline
40&$\bar{u}_{R}^{\bar{c1}}$&$\stackrel{03}{[-i]}\,\stackrel{12}{[-]}|\stackrel{56}{[-]}\,\stackrel{78}{(+)}
||\stackrel{9 \;10}{[-]}\;\;\stackrel{11\;12}{[+]}\;\;\stackrel{13\;14}{[+]}$&1&$-\frac{1}{2}$&-1&
$-\frac{1}{2}$&0&$-\frac{1}{2}$&$-\frac{1}{2\,\sqrt{3}}$&$-\frac{1}{6}$&$-\frac{1}{6}$&$-\frac{2}{3}$\\
\hline\hline
41&$ \bar{d}_{L}^{\bar{c2}}$&$ \stackrel{03}{[-i]}\,\stackrel{12}{(+)}|
\stackrel{56}{(+)}\,\stackrel{78}{(+)}
||\stackrel{9 \;10}{(+)}\;\;\stackrel{11\;12}{(-)}\;\;\stackrel{13\;14}{[+]} $ &-1&$\frac{1}{2}$&1&0&
$\frac{1}{2}$&$\frac{1}{2}$&$-\frac{1}{2\,\sqrt{3}}$&$-\frac{1}{6}$&$\frac{1}{3}$&$\frac{1}{3}$\\
\hline 
$\cdots$ &&&&&&&&&&&& \\
\hline\hline
49&$ \bar{d}_{L}^{\bar{c3}}$&$ \stackrel{03}{[-i]}\,\stackrel{12}{(+)}|
\stackrel{56}{(+)}\,\stackrel{78}{(+)}
||\stackrel{9 \;10}{(+)}\;\;\stackrel{11\;12}{[+]}\;\;\stackrel{13\;14}{(-)} $ &-1&$\frac{1}{2}$&1&0&
$\frac{1}{2}$&$0$&$\frac{1}{\sqrt{3}}$&$-\frac{1}{6}$&$\frac{1}{3}$&$\frac{1}{3}$\\
\hline 
$\cdots$ &&&&&&&&&&&& \\
\hline\hline
57&$ \bar{e}_{L}$&$ \stackrel{03}{[-i]}\,\stackrel{12}{(+)}|
\stackrel{56}{(+)}\,\stackrel{78}{(+)}
||\stackrel{9 \;10}{[-]}\;\;\stackrel{11\;12}{(-)}\;\;\stackrel{13\;14}{(-)} $ &-1&$\frac{1}{2}$&1&0&
$\frac{1}{2}$&$0$&$0$&$\frac{1}{2}$&$1$&$1$\\
\hline 
58&$\bar{e}_{L}$&$\stackrel{03}{(+i)}\,\stackrel{12}{[-]}|\stackrel{56}{(+)}\,\stackrel{78}{(+)}
||\stackrel{9 \;10}{[-]}\;\;\stackrel{11\;12}{(-)}\;\;\stackrel{13\;14}{(-)}$&-1&$-\frac{1}{2}$&1&0&
$\frac{1}{2}$ &$0$&$0$&$\frac{1}{2}$&$1$&$1$\\
\hline
59&$\bar{\nu}_{L}$&$\stackrel{03}{[-i]}\,\stackrel{12}{(+)}|\stackrel{56}{[-]}\,\stackrel{78}{[-]}
||\stackrel{9 \;10}{[-]}\;\;\stackrel{11\;12}{(-)}\;\;\stackrel{13\;14}{(-)}$&-1&$\frac{1}{2}$&1&0&
$-\frac{1}{2}$&$0$&$0$&$\frac{1}{2}$&$0$&$0$\\
\hline 
60&$ \bar{\nu}_{L} $&$\stackrel{03}{(+i)}\,\stackrel{12}{[-]}|
\stackrel{56}{[-]}\,\stackrel{78}{[-]}
||\stackrel{9 \;10}{[-]}\;\;\stackrel{11\;12}{(-)}\;\;\stackrel{13\;14}{(-)} $&-1&$-\frac{1}{2}$&1&0&
$-\frac{1}{2}$&$0$&$0$&$\frac{1}{2}$&$0$&$0$\\
\hline
61&$\bar{\nu}_{R}$&$\stackrel{03}{(+i)}\,\stackrel{12}{(+)}|\stackrel{56}{[-]}\,\stackrel{78}{(+)}
||\stackrel{9 \;10}{[-]}\;\;\stackrel{11\;12}{(-)}\;\;\stackrel{13\;14}{(-)}$&1&$\frac{1}{2}$&-1&
$-\frac{1}{2}$&0&$0$&$0$&$\frac{1}{2}$&$\frac{1}{2}$&$0$\\
\hline
62&$\bar{\nu}_{R} $&$\stackrel{03}{[-i]}\,\stackrel{12}{[-]}|\stackrel{56}{[-]}\,\stackrel{78}{(+)}
||\stackrel{9 \;10}{[-]}\;\;\stackrel{11\;12}{(-)}\;\;\stackrel{13\;14}{(-)} $&1&$-\frac{1}{2}$&-1&
$-\frac{1}{2}$&0&$0$&$0$&$\frac{1}{2}$&$\frac{1}{2}$&$0$\\
\hline
63&$ \bar{e}_{R}$&$\stackrel{03}{(+i)}\,\stackrel{12}{(+)}|\stackrel{56}{(+)}\,\stackrel{78}{[-]}
||\stackrel{9 \;10}{[-]}\;\;\stackrel{11\;12}{(-)}\;\;\stackrel{13\;14}{(-)}$ &1&$\frac{1}{2}$&-1&
$\frac{1}{2}$&0 &$0$&$0$&$\frac{1}{2}$&$\frac{1}{2}$&$1$\\
\hline
64&$\bar{e}_{R}$&$\stackrel{03}{[-i]}\,\stackrel{12}{[-]}|\stackrel{56}{(+)}\,\stackrel{78}{[-]}
||\stackrel{9 \;10}{[-]}\;\;\stackrel{11\;12}{(-)}\;\;\stackrel{13\;14}{(-)}$&1&$-\frac{1}{2}$&-1&
$\frac{1}{2}$&0&$0$&$0$&$\frac{1}{2}$&$\frac{1}{2}$&$1$\\
\hline 
\end{tabular}}
\label{Table so13+1.b}
\end{sidewaystable}
%}
%
%\end{center}
%\end{small}
%\end{tiny}
%

Since spinors (fermions) carry besides the family members quantum numbers also the family 
quantum numbers, determined by $\tilde{S}^{ab}= \frac{i}{4} (\tilde{\gamma}^{a} 
\tilde{\gamma}^{b} - \tilde{\gamma}^{b}\tilde{\gamma}^{a})$, there are correspondingly
$2^{(7+1)/2-1} =8$ families~\cite{JMP2015}, which split into two groups of families, each
 manifesting the ($\widetilde{SU}(2)_{\widetilde{SO}(3,1)}$ $\times \widetilde{SU}(2)_{\widetilde{SO}(4)}$
$\times U(1)$) symmetry.

%%%%%%%%%%%%%%%%%%%%%%%%%%%
The eight families of the first member of the eight-plet of quarks from Tables~\ref{Table so13+1.}%
--\ref{Table so13+1.b}, 
for example, that is of the right  handed $u_{1R}$ quark,  are 
presented in the left column of  Table~\ref{Table III.}~\cite{JMP}. In the right column of the 
same table the equivalent eight-plet of the right handed neutrinos $\nu_{1R}$ are presented.
All the other members of any of the eight families of quarks or leptons follow  from any member 
of a particular family by the application of the  operators $N^{\pm}_{R,L}$ and  $\tau^{(2,1)\pm}$ 
on this particular member.  

%\clearpage
 
%
 \begin{sidewaystable}
%\begin{tiny}
\tbl{
Eight families of the right handed $u^{c1}_{R}$ (\ref{Table so13+1.}--\ref{Table so13+1.b}) 
quark with spin $\frac{1}{2}$, the colour charge $(\tau^{33}=1/2$, $\tau^{38}=1/(2\sqrt{3})$, 
and of  the colourless right handed neutrino $\nu_{R}$ of spin $\frac{1}{2}$ %(\ref{Table II.})  
are presented in the  left and in the right column, respectively.
They belong to two groups of four families, one ($I$) is a doublet with respect to 
($\vec{\tilde{N}}_{L}$ and  $\vec{\tilde{\tau}}^{(1)}$) and  a singlet with respect to 
($\vec{\tilde{N}}_{R}$ and  $\vec{\tilde{\tau}}^{(2)}$), the other ($II$) is a singlet with respect to 
($\vec{\tilde{N}}_{L}$ and  $\vec{\tilde{\tau}}^{(1)}$) and  a doublet with with respect to 
($\vec{\tilde{N}}_{R}$ and  $\vec{\tilde{\tau}}^{(2)}$).
All the families follow from the starting one by the application of the operators 
($\tilde{N}^{\pm}_{R,L}$, $\tilde{\tau}^{(2,1)\pm}$), Eq.~(\ref{plusminus}).  The generators 
($N^{\pm}_{R,L} $, $\tau^{(2,1)\pm}$) (Eq.~(\ref{plusminus}))
transform $u_{1R}$ to all the members of one family of the same colour. 
The same generators transform equivalently the right handed   neutrino $\nu_{1R}$  to all the colourless 
members of the same family.
}
 %\begin{center}
 {\begin{tabular}{r|c|c|c|c|c c c c c}
 \hline
 &&&&&$\tilde{\tau}^{13}$&$\tilde{\tau}^{23}$&$\tilde{N}_{L}^{3}$&$\tilde{N}_{R}^{3}$&$\tilde{\tau}^{4}$\\
 \hline
 $I$&$u^{c1}_{R\,1}$&
   $ \stackrel{03}{(+i)}\,\stackrel{12}{[+]}|\stackrel{56}{[+]}\,\stackrel{78}{(+)} ||
   \stackrel{9 \;10}{(+)}\;\;\stackrel{11\;12}{(-)}\;\;\stackrel{13\;14}{(-)}$ & 
   $\nu_{R\,2}$&
   $ \stackrel{03}{(+i)}\,\stackrel{12}{[+]}|\stackrel{56}{[+]}\,\stackrel{78}{(+)} ||
   \stackrel{9 \;10}{(+)}\;\;\stackrel{11\;12}{[+]}\;\;\stackrel{13\;14}{[+]}$ 
  &$-\frac{1}{2}$&$0$&$-\frac{1}{2}$&$0$&$-\frac{1}{2}$ 
 \\
  $I$&$u^{c1}_{R\,2}$&
   $ \stackrel{03}{[+i]}\,\stackrel{12}{(+)}|\stackrel{56}{[+]}\,\stackrel{78}{(+)} ||
   \stackrel{9 \;10}{(+)}\;\;\stackrel{11\;12}{(-)}\;\;\stackrel{13\;14}{(-)}$ & 
   $\nu_{R\,2}$&
   $ \stackrel{03}{[+i]}\,\stackrel{12}{(+)}|\stackrel{56}{[+]}\,\stackrel{78}{(+)} ||
   \stackrel{9 \;10}{(+)}\;\;\stackrel{11\;12}{[+]}\;\;\stackrel{13\;14}{[+]}$ 
  &$-\frac{1}{2}$&$0$&$\frac{1}{2}$&$0$&$-\frac{1}{2}$
 \\
  $I$&$u^{c1}_{R\,3}$&
   $ \stackrel{03}{(+i)}\,\stackrel{12}{[+]}|\stackrel{56}{(+)}\,\stackrel{78}{[+]} ||
   \stackrel{9 \;10}{(+)}\;\;\stackrel{11\;12}{(-)}\;\;\stackrel{13\;14}{(-)}$ & 
   $\nu_{R\,3}$&
   $ \stackrel{03}{(+i)}\,\stackrel{12}{[+]}|\stackrel{56}{(+)}\,\stackrel{78}{[+]} ||
   \stackrel{9 \;10}{(+)}\;\;\stackrel{11\;12}{[+]}\;\;\stackrel{13\;14}{[+]}$ 
  &$\frac{1}{2}$&$0$&$-\frac{1}{2}$&$0$&$-\frac{1}{2}$
 \\
 $I$&$u^{c1}_{R\,4}$&
  $ \stackrel{03}{[+i]}\,\stackrel{12}{(+)}|\stackrel{56}{(+)}\,\stackrel{78}{[+]} ||
  \stackrel{9 \;10}{(+)}\;\;\stackrel{11\;12}{(-)}\;\;\stackrel{13\;14}{(-)}$ & 
  $\nu_{R\,4}$&
  $ \stackrel{03}{[+i]}\,\stackrel{12}{(+)}|\stackrel{56}{(+)}\,\stackrel{78}{[+]} ||
  \stackrel{9 \;10}{(+)}\;\;\stackrel{11\;12}{[+]}\;\;\stackrel{13\;14}{[+]}$ 
  &$\frac{1}{2}$&$0$&$\frac{1}{2}$&$0$&$-\frac{1}{2}$
  \\
  \hline
  $II$& $u^{c1}_{R\,5}$&
        $ \stackrel{03}{[+i]}\,\stackrel{12}{[+]}|\stackrel{56}{[+]}\,\stackrel{78}{[+]}||
        \stackrel{9 \;10}{(+)}\;\;\stackrel{11\;12}{(-)}\;\;\stackrel{13\;14}{(-)}$ & 
        $\nu_{R\,5}$&
        $ \stackrel{03}{[+i]}\,\stackrel{12}{[+]}|\stackrel{56}{[+]}\,\stackrel{78}{[+]}|| 
        \stackrel{9 \;10}{(+)}\;\;\stackrel{11\;12}{[+]}\;\;\stackrel{13\;14}{[+]}$ 
        &$0$&$-\frac{1}{2}$&$0$&$-\frac{1}{2}$&$-\frac{1}{2}$
 \\ 
  $II$& $u^{c1}_{R\,6}$&
      $ \stackrel{03}{(+i)}\,\stackrel{12}{(+)}|\stackrel{56}{[+]}\,\stackrel{78}{[+]}||
      \stackrel{9 \;10}{(+)}\;\;\stackrel{11\;12}{(-)}\;\;\stackrel{13\;14}{(-)}$ & 
      $\nu_{R\,6}$&
      $ \stackrel{03}{(+i)}\,\stackrel{12}{(+)}|\stackrel{56}{[+]}\,\stackrel{78}{[+]}|| 
      \stackrel{9 \;10}{(+)}\;\;\stackrel{11\;12}{[+]}\;\;\stackrel{13\;14}{[+]}$ 
      &$0$&$-\frac{1}{2}$&$0$&$\frac{1}{2}$&$-\frac{1}{2}$
 \\ 
 $II$& $u^{c1}_{R\,7}$&
 $ \stackrel{03}{[+i]}\,\stackrel{12}{[+]}|\stackrel{56}{(+)}\,\stackrel{78}{(+)}||
 \stackrel{9 \;10}{(+)}\;\;\stackrel{11\;12}{(-)}\;\;\stackrel{13\;14}{(-)}$ & 
      $\nu_{R\,7}$&
      $ \stackrel{03}{[+i]}\,\stackrel{12}{[+]}|\stackrel{56}{(+)}\,\stackrel{78}{(+)}|| 
      \stackrel{9 \;10}{(+)}\;\;\stackrel{11\;12}{[+]}\;\;\stackrel{13\;14}{[+]}$ 
    &$0$&$\frac{1}{2}$&$0$&$-\frac{1}{2}$&$-\frac{1}{2}$
  \\
   $II$& $u^{c1}_{R\,8}$&
    $ \stackrel{03}{(+i)}\,\stackrel{12}{(+)}|\stackrel{56}{(+)}\,\stackrel{78}{(+)}||
    \stackrel{9 \;10}{(+)}\;\;\stackrel{11\;12}{(-)}\;\;\stackrel{13\;14}{(-)}$ & 
    $\nu_{R\,8}$&
    $ \stackrel{03}{(+i)}\,\stackrel{12}{(+)}|\stackrel{56}{(+)}\,\stackrel{78}{(+)}|| 
    \stackrel{9 \;10}{(+)}\;\;\stackrel{11\;12}{[+]}\;\;\stackrel{13\;14}{[+]}$ 
    &$0$&$\frac{1}{2}$&$0$&$\frac{1}{2}$&$-\frac{1}{3}$
 \\ 
 \hline 
 \end{tabular}
}
% \end{center}
%\end{tiny}
%
\label{Table III.} 
 \end{sidewaystable}
The eight-plets separate into two group of four families: One group  contains  doublets with respect 
to $\vec{\tilde{N}}_{R}$ and  $\vec{\tilde{\tau}}^{2}$, these families are singlets with respect to 
$\vec{\tilde{N}}_{L}$ and  $\vec{\tilde{\tau}}^{1}$ (with the eigenvalues $0$). Another group of 
families contains  doublets 
with respect to  $\vec{\tilde{N}}_{L}$ and  $\vec{\tilde{\tau}}^{1}$, these families are singlets 
with respect to  $\vec{\tilde{N}}_{R}$ and  $\vec{\tilde{\tau}}^{2}$ (with the eigenvalues $0$). 

The scalar fields, which are the gauge scalars  of  $\vec{\tilde{N}}_{R}$ and  
$\vec{\tilde{\tau}}^{2}$,  
couple only to the four families  which are doublets with respect to these two groups. 
The scalar fields, which are the gauge scalars  of  $\vec{\tilde{N}}_{L}$ and  
$\vec{\tilde{\tau}}^{1}$, 
couple only to the four families  which are doublets with respect to these last two groups.

 If there are no fermions present then the  vector gauge fields of the family members charges 
and of the family charges -  $\omega_{abm}$ and $\tilde{\omega}_{abm}$, respectively - are 
uniquely expressible with the vielbeins~\cite{norma2014MatterAntimatter,JMP2015}.
%, which are then the only propagating fields. 

The scalar fields, the gauge fields with the space index  $s=(7,8)$, which are either 
superposition of $\tilde{\omega}_{abs}$ or  of $\omega_{s'ts}$, determine - 
after gaining nonzero vacuum expectation values (the assumption {\bf A v.} and comments 
{\bf C v.}) - masses of fermions (belonging to two groups of four families of family members of 
spinors)   and weak bosons.

The condensate (the assumption {\bf A iv.}), Table~\ref{Table con.}, gives masses of the order
of the scale of its appearance to
% $A^{Y'}_{m}=  \cos \vartheta_{2} A^{23}_{m} - \sin \vartheta_{2} A^{4}_{m}$, and to 
all the scalar gauge fields, presented in the second and the third line of Eq.~(\ref{faction}).
%, leading to $A^{Ai}_{s}, s\in (5,6,\dots,13,14)$ and $\tilde{A}^{Ai}_{t}$, 
%$t\in (5,6,\dots,13,14)$.  

%%%%%%%%%%%%%%%%%%%%%%%%%% Friday, 13.05.2016 at 16:06

The colour, the weak and the hyper charges  ($\vec{\tau}^{3}$, $\vec{\tau^{1}}, Y$,
respectively) of the corresponding gauge fields are  before the 
electroweak break the conserved charges, since the corresponding vector gauge fields don't  
interact with  the condensate. After the electroweak break, when the scalar fields with the space
index $s=(7,8)$ - those with the family quantum numbers and those with the quantum numbers
($Q,Q',Y'$) - start to strongly self interact~(Eq.~(\ref{interactingphi})), gaining nonzero vacuum
expectation values, the weak charge and the hyper charge are no longer conserved. The only 
conserved charges are then the colour and the electromagnetic charges.

In Eq.~(\ref{checktildeNL3Q}) (the reader may have a look at the Ref.~\cite{JMP2015},
Eqs.~(10, A8, A9)) the scalar fields with the space index $(7,8)$, 
Eq.~(\ref{commonAi}), are  presented as superposition of the spin connection fields of both 
kinds. These scalar fields determine after the electroweak break the mass matrices of the two 
decoupled groups of four families (Eq.~(\ref{M0})) and of the heavy bosons.
 
%Getting nonzero vacuum expectation values they 
%cause the electroweak break, changing also their own masses. These scalar fields determine also the masses 
%of the gauge bosons.
%%%%%%%%%%%%%%%%%%%%%%%%%%%%%%%%%%%%%

%%%%%%%%%%%%%%%%%%%%%%%%%%%%%%

%Quarks and leptons have the "spinor" quantum number ($\tau^{4}$, originating in $SO(6)$  
%(Eq.~(\ref{so64}), presented in Tables~\ref{Table so13+1.}--\ref{Table so13+1.b}) equal to 
%$\frac{1}{6}$ and 
%$-\frac{1}{2}$, respectively~\footnote{In the Pati-Salam 
%model~\cite{patisalam} this "spinor" quantum number is named $\frac{B-L}{2}$ quantum 
%number and is twice the "spinor" quantum number, for quarks equal to $\frac{1}{3}$ and for 
%leptons to $-1$.} (with the sum of both equal to $3 \times \frac{1}{6}+ (-\frac{1}{2})=0$). 

%% Manjka:
% A.  Matter/atimatter asymmetry
% B.  Dark Matter
% C.  Fourth family and two photons

%Is this what follows written already above? CHECK****

%
\section{The scalar fields contributing to the electroweak break belong to the weak charge 
doublets}
\label{CPNandpropScm}

This section follows mainly the equivalent sections in Refs.~\cite{norma2014MatterAntimatter,%
JMP2015}.

It turns out~\cite{norma2014MatterAntimatter} that all scalars (the gauge fields with the space
index $s \ge 5$) of the action (Eq.~\ref{action}) carry with respect to the space index charges in 
the fundamental representations: They are either doublets (Table~\ref{Table doublets.}), 
$s=(5,6,7,8)$, or triplets (Ref.~\cite{norma2014MatterAntimatter}, Sect. II, Table I), 
$s=(9,10,..,13,14)$. 
The scalars with the space indices $s \in (7,8)$ and $s \in (5,6)$ are 
%, with respect to this degree of freedom, 
the $SU(2)$ {\em doublets} (Table~\ref{Table doublets.}). 

It is demonstrated in this section that all the scalar gauge fields with the space index $s \in (7,8)$ 
carry - with respect to the space index $s \in (7,8)$ - the weak and the hyper charge as does 
the Higgs's scalar of the {\it standard model} ($\pm (\frac{1}{2}, -\frac{1}{2})$), while they 
carry in addition either the family quantum numbers, belonging to (one of the two times two
$\widetilde{SU}(2)$) triplets, or the family members quantum numbers, belonging to 
(one of the three) singlets. These scalars offer the 
explanation for the origin of the Higgs's scalar and the Yukawa coupling (of the scalar higgs to 
fermions) of the {\it standard model}.

To see this one must take into account %(Eq.~\ref{sabtildesab})
that the infinitesimal generators $S^{ab}$, 
\begin{eqnarray}
\label{sabsc0}
S^{ab} &=& \,\frac{i}{4} (\gamma^a\, \gamma^b - \gamma^b\, \gamma^a)\,,
\end{eqnarray}
determine spins of spinors%(Eq.~(\ref{sabsc}))
, while  $\tilde{S}^{ab}$,  
\begin{eqnarray}
\label{tildesabsc0}
\tilde{S}^{ab} &=& \,\frac{i}{4} (\tilde{\gamma}^a\, \tilde{\gamma}^b - 
\tilde{\gamma}^b\, \tilde{\gamma}^a)\,,
\end{eqnarray}
determine family charges of spinors (Eq.~(\ref{tildesabsc0})), while
${\cal S}^{ab}$~(Assumption {\bf A i.}%(Eq.\ref{gammatildedep}
), 
which apply on the spin connections $\omega_{bd e}$ ($= f^{\alpha}{}_{e}\, $ 
$\omega_{bd \alpha}$) and $\tilde{\omega}_{\tilde{b} \tilde{d} e}$ ($= f^{\alpha}{}_{e}\,$ 
$\tilde{\omega}_{\tilde{b} \tilde{d} \alpha}$), on either the space index $e$ or any of the
indices $(b,d,\tilde{b},\tilde{d})$, operates as follows 
\begin{eqnarray}
\label{bosonspin0}
{\cal S}^{ab} \, A^{d\dots e \dots g} &=& i \,(\eta^{ae} \,A^{d\dots b \dots g} - 
\eta^{be}\,A^{d\dots a \dots g} )\,,
\end{eqnarray}
in accordance with the Eqs.~(\ref{grapheigen}, \ref{snmb:gammatildegamma}, 
\ref{stildestrans}).
The expressions for the infinitesimal operators of the subgroups of the starting groups 
(presented in Eq.~(\ref{tau}) and the footnote before this  Eq.~(\ref{tau}) and determined by 
the coefficients %- %\ref{so42}, \ref{so64}, \ref{so1+3tilde}, \ref{so42tilde}, \ref{so64tilde}, 
%\ref{so1+3boson}, \ref{so42boson}, \ref{so64boson}, %\ref{YQY'Q'andtilde})) are equivalent (have for the chosen $Ai$ the same coefficients
$c^{Ai}{}_{ab}$ in Eq.~(\ref{tau}))  are the same for all three kinds of degrees of freedom 
%(Apps.~\ref{thetagammatildegammaappendix}, 
%\ref{generatorsandfields}).
(Eqs.~(\ref{sabsc0}, \ref{tildesabsc0}, \ref{bosonspin0})).

All scalars carry correspondingly, besides the quantum numbers determined by the space index
$s$, also the quantum numbers $Ai$, Eq.~(\ref{tau}), the states of which belong to the adjoint
representations. 
%They namely couple to fermions also through the family (carrying $\tilde{\tau}^{\tilde{A}i}$) 
%and family members (carrying $\tau^{Ai}$) quantum numbers.
 At the electroweak break all the scalar fields with the space index $(7,8)$, those which belong
to one of twice two triplets carrying the family quantum numbers ($\tilde{\tau}^{\tilde{A}i}$) 
%Eqs.~(\ref{so1+3tilde}- \ref{so64tilde}))
 and those which belong to one of the three singlets carrying the family members quantum
numbers ($Q,Q',Y'$), Eq.~(\ref{commonAi}) 
%, Sect.~\ref{theaction}, the assumption {\bf v.} and the corresponding comments), 
start to self interact, gaining nonzero vacuum expectation values and breaking the weak charge, 
the hyper charge and the family charges.

Let me introduce  a  common notation  $ A^{Ai}_{s}$ for all the scalar fields with $s=(7,8)$,
independently of whether  they originate in $\omega_{abs}$ - in this case $ Ai$ 
$=(Q$,$Q',Y'$) - or in $\tilde{\omega}_{\tilde{a}\tilde{b}s}$ - in this case all the
family quantum numbers of all eight families contribute. 
\begin{eqnarray}
\label{commonAi}
 A^{Ai}_{s} &{\rm represents}& (\,A^{Q}_{s}\,,A^{Q'}_{s}\,, A^{Y'}_{s}\,, 
 % \tilde{\omega}_{\tilde{a} \tilde{b} s }\,,
 \vec{\tilde{A}}^{\tilde{1}}_{s}\,, 
 \vec{\tilde{A}}^{\tilde{N}_{\tilde{L}}}_{s}\,, \vec{\tilde{A}}^{\tilde{2}}_{s}\,, 
 \vec{\tilde{A}}^{\tilde{N}_{\tilde{R}}}_{s}\,)\,,\nonumber\\
\tau^{Ai} &{\rm represents}& (Q,\,Q',\,Y', \,\vec{\tilde{\tau}}^{1},\, \vec{\tilde{N}}_{L},\,
\vec{\tilde{\tau}}^{2},\,\vec{\tilde{N}}_{R})\,.
\end{eqnarray}
Here $\tau^{Ai}$ represent all the operators, which apply on the spinor states.
These scalars, the gauge scalar fields of the generators $\tau^{Ai}$ and $\tilde{\tau}^{Ai}$ 
%(Eqs.~(\ref{so64} - \ref{so42tilde})), 
are expressible in terms of the spin connection fields (Ref.~\cite{JMP2015}, 
Eqs.~(10,22,A8,A9).

{\it Statement 1:} Scalar fields with the space index $(7,8)$ carry with respect to this space
index the weak and the hyper charge ($\mp \frac{1}{2}$, $\pm \frac{1}{2}$), respectively.

The proof is presented in Ref.~\cite{JMP2015}. I shall here demonstrate it only briefly.

Let us make a choice of the superposition of the scalar fields so that they are eigenstates of 
$\tau^{13}= \frac{1}{2}({\cal S}^{56} - {\cal S}^{78})$ (Eq~(\ref{tau}) and footnotes 
at the same page). For this purpose let us rewrite the second line of Eq.~(\ref{faction}) 
as follows (the momentum $p_{s}$ is left out~\footnote{It is expected that solutions with 
nonzero momentum lead to higher masses of the fermion fields in $d=(3+1)$~\cite{DHN,DN012}.
We correspondingly pay no attention to the momentum $p_{s}\,, s\in(4,8)$, when having in mind
the lowest energy solutions, manifesting at low energies.})
\begin{eqnarray}
\label{eigentau1tau2}
 & &\sum_{s=(7,8), Ai}\, \bar{\psi} \,\gamma^s\, ( - \tau^{Ai} \,A^{Ai}_{s}\,)\,\psi =
\nonumber\\
 & &- \bar{\psi}\,\{\,\stackrel{78}{(+)}\, \tau^{Ai} \,(A^{Ai}_{7} - i   
 \,A^{Ai}_{8})\, + \stackrel{78}{(-)}(\tau^{Ai} \,(A^{Ai}_{7} + i \,A^{Ai}_{8})\,\}\,\psi\,,
 \nonumber\\
 & &\stackrel{78}{(\pm)} = \frac{1}{2}\, (\gamma^{7} \pm \,i \, \gamma^{8}\,)\,,\quad
 A^{Ai}_{\scriptscriptstyle{\stackrel{78}{(\pm)}}}: = (A^{Ai}_7 \,\mp i\, A^{Ai}_8)\,,
\end{eqnarray}
with the summation over $Ai$ performed, since $A^{Ai}_s$ represent the scalar fields 
($A^{Q}_{s}$, $A^{Q'}_{s}$, $A^{Y'}_{s}$, 
$\tilde{A}^{\tilde{4}}_{s}$, %$\tilde{A}^{\tilde{Q}}_{s}$, 
$\vec{\tilde{A}}^{\tilde{1}}_{s}$, $\vec{\tilde{A}}^{\tilde{2}}_{s}$,
 $\vec{\tilde{A}}^{\tilde{N}_{R}}_{s}$ and  $\vec{\tilde{A}}^{\tilde{N}_{L}}_{s}$).

The application of the operators $Y$ ($Y= \tau^{23} +\tau^{4}$, $\tau^{23} =
 \frac{1}{2} ({\cal S}^{56} +{\cal S}^{78})$, $\tau^{4}=$ $ -\frac{1}{3} ({\cal S}^{9\,10}
 +  {\cal S}^{11\,12} +{\cal S}^{13\,14})$), $Q$ ($Q= \tau^{13} +Y $ 
and  $ \tau^{13}$ ($\tau^{13} = \frac{1}{2} ({\cal S}^{56} - {\cal S}^{78})$)  
on the fields ($A^{Ai}_{7}\mp i\,A^{Ai}_{8})$ gives  (${\cal S}^{ab}$ is defined in 
Eq.~(\ref{bosonspin0})) 

\begin{eqnarray}
\label{checktau13Y}
\tau^{13}\,(A^{Ai}_7 \,\mp i\, A^{Ai}_8)&=& \pm \,\frac{1}{2}\,(A^{Ai}_7 \,
\mp i\, A^{Ai}_8)\,,\nonumber\\
Y\,(A^{Ai}_7 \,\mp i\, A^{Ai}_8)&=& \mp \,\frac{1}{2}\,(A^{Ai}_7 \,
\mp i\, A^{Ai}_8)\,,\nonumber\\
Q\,(A^{Ai}_7 \,\mp i\, A^{Ai}_8)&=& 0\,.
\end{eqnarray}
Since  $ \tau^{4}$, $Y$, $\tau^{13}$ and %$\tau^{1\splus},  \tau^{1\sminus}$ this is correct,
% not the following two
$\tau^{1 +},  \tau^{1 -}$ give zero if 
applied on  ($A^{Q}_{s}$, $A^{Q'}_{s}$ and $A^{Y'}_{s}$) with  respect to the indices 
($Q,Q',Y'$),  and since $Y$ and $\tau^{13}$ commute with the family quantum numbers, one 
sees that the scalar fields $A^{Ai}_{s}$ ( =$A^{Q}_{s}$, $A^{Y}_{s}$, $A^{Y'}_{s}$, 
$\tilde{A}^{\tilde{4}}_{s}$, $\tilde{A}^{\tilde{Q}}_{s}$, $\vec{\tilde{A}}^{\tilde{1}}_{s}$, 
$\vec{\tilde{A}}^{\tilde{2}}_{s}$, $\vec{\tilde{A}}^{\tilde{N}_{R}}_{s}$, 
$\vec{\tilde{A}}^{\tilde{N}_{L}}_{s}$), rewritten as  %
%\begin{eqnarray}
%\label{Ytau13}
$A^{Ai}_{\scriptscriptstyle{\stackrel{78}{(\pm)}}} $ $= (A^{Ai}_7 \,\mp i\, A^{Ai}_8)\,$,
%\end{eqnarray}
%
are eigenstates of $\tau^{13}$ and $Y$, having the quantum numbers of the {\it standard model} 
Higgs' scalar.
%~\footnote{$\tau^{Ai}$, $Y$ and $Q$ determine operators which apply on scalar fields.}. 

These superposition of $A^{Ai}_{\scriptscriptstyle{\stackrel{78}{(\pm)}}}$ are presented in 
Table~\ref{Table doublets.} as two doublets with respect to the weak charge 
${\cal \tau}^{13}$,  with the eigenvalue of ${\cal \tau}^{23}$  (the second 
$SU(2)_{II}$ charge), 
equal to either $-\frac{1}{2}$ or $+\frac{1}{2}$, respectively. 
\begin{table}
\tbl{%\label{Table doublets.} 
The two scalar weak doublets, one with $ {\cal \tau}^{23}=- \frac{1}{2}$  and the other with 
$ {\cal \tau}^{23}=+ \frac{1}{2}$, both with the "spinor" quantum number ${\cal \tau}^{4}$ 
$=0$, are 
presented.
In this table all the scalar fields carry besides the quantum numbers determined by the space index 
also  the quantum numbers ${\cal A}i$ from Eq.~(\ref{commonAi}).}
% \begin{center}
 %\begin{small}
{ \begin{tabular}{c|c| c c c c r}
 \hline
 &state & ${\cal \tau}^{13}$& $ {\cal \tau}^{23}$ & spin& ${\cal \tau}^{4}$& $ Q$\\
 \hline
 $A^{Ai}_{\scriptscriptstyle{\stackrel{78}{(-)}}}$ & $A^{Ai}_{7}+iA^{Ai}_{8}$& $+
\frac{1}{2}$& 
 $-\frac{1}{2}$& 0&0& 0\\
 $A^{Ai}_{\scriptscriptstyle{\stackrel{56}{(-)}}}$ & $A^{Ai}_{5}+iA^{Ai}_{6}$& 
$-\frac{1}{2}$& 
 $-\frac{1}{2}$& 0&0& -1\\
 \hline 
$A^{Ai}_{\scriptscriptstyle{\stackrel{78}{(+)}}}$ & $A^{Ai}_{7}-iA^{Ai}_{8}$& 
$-\frac{1}{2}$& 
$+\frac{1}{2}$& 0&0& 0\\
$A^{Ai}_{\scriptscriptstyle{\stackrel{56}{(+)}}}$ & $A^{Ai}_{5}-iA^{Ai}_{6}$& $+
\frac{1}{2}$& 
$+\frac{1}{2}$& 0& 0&+1\\ 
\hline
\end{tabular}
}
 %\end{small}
% \end{center}
\label{Table doublets.}
 \end{table}

The operators ${\cal \tau}^{1\spm} = {\cal \tau}^{11}\pm i {\cal \tau}^{12} $
\begin{eqnarray}
\label{YQ}
\tau^{1\spm} &=& \frac{1}{2} [({\cal S}^{58}- {\cal S}^{67})\,\smp \,i\,
%\tau^{1\pm} &=& \frac{1}{2} [({\cal S}^{58}- {\cal S}^{67})\,\mp \,i\,
({\cal S}^{57}+ {\cal S}^{68})]\,, % \tau^{13} = \frac{1}{2} (S^{56}- S^{78})\,,\nonumber\\
%Y&=& \tau^{23} + \tau^{4}\,,\quad Q= Y+\tau^{13}\,, \nonumber
\end{eqnarray}
transform one member of a doublet from Table~\ref{Table doublets.} into another member of 
the same doublet, keeping $ \tau^{23}$  ($= \frac{1}{2}\,({\cal S}^{56}+ {\cal S}^{78})$) 
unchanged, clarifying the above statement.

After the appearance of the condensate (Table~\ref{Table con.}), which breaks the
 $SU(2)_{II}$ symmetry (bringing masses to all the scalar fields), the weak charge, 
$\vec{\tau}^{1}$, and the hyper charge $Y$ remain the conserved
 charges~\footnote{It is $\tau^{23}$ which determines the hyper charge $Y$ 
($ Y= {\cal S}^{23} + \tau^{4}$) of these scalar fields, since $\tau^{4}$, if applied on the 
scalar index of these scalar fields, gives zero, according to Eqs. in the footnote above 
Eq.~(\ref{tau}).}.

At the electroweak break the scalar fields with the space index $(7,8)$ start to interact among 
themselves so that the Lagrange density for these gauge fields changes from
${\cal L}_{s} =E\, \{(p_{m} \,\Phi^{Ai}_{s})^{\dagger} \, (p^{m} \,\Phi^{Ai}_{s}) -
(m'_{Ai})^2 \, \Phi^{Ai\dagger}_{s} \,\Phi^{Ai}_{s}\}$ to 
\begin{eqnarray}
\label{interactingphi}
{\cal L}_{sg} &=& E\,\sum_{A,i} \,\{(p_{m} \,\Phi^{Ai}_{s})^{\dagger} \,(p^{m} \,
\Phi^{Ai}_{s}) - (-\lambda^{Ai} + (m'_{Ai})^2)) \,\Phi^{Ai \dagger}_{s} \Phi^{Ai}_{s} 
\nonumber\\ 
&+& \sum_{B,j}\, \Lambda^{Ai Bj}\,\Phi^{Ai \dagger}_{s} \Phi^{Ai}_{s} \;
\Phi^{Bj \dagger}_{s} \Phi^{Bj}_{s}\}\,,
\end{eqnarray}
where $-\lambda^{Ai} +  m'^{2}_{Ai}= m^{2}_{Ai}$  and $m_{Ai}$ manifests as the mass of
the $A^{Ai}_{s}$ scalar.

The operator 
${\cal \tau}^{1\splus}$ (Eq.~(\ref{YQ}))
 transforms 
$A^{Ai}_{\scriptscriptstyle{\stackrel{78}{(\pm)}}}$ into
 $A^{Ai}_{\scriptscriptstyle{\stackrel{56}{(\pm)}}}:=$ $(A^{Ai}_5 \,\mp i\, A^{Ai}_6)$,
 while
${\cal \tau}^{1\sminus} \,A^{Ai}_{\scriptscriptstyle{\stackrel{78}{(\pm)}}} =0.$
%\begin{eqnarray}
%\label{tau1pm}
%\tau^{1\splus} \,A^{Ai}_{\scriptscriptstyle{\stackrel{78}{(\pm)}}} &=& (A^{Ai}_5 \,\mp i\, A^{Ai}_6)= 
%:A^{Ai}_{\scriptscriptstyle{\stackrel{56}{(\pm)}}}\,,
%\nonumber\\
%\tau^{1\sminus} \,A^{Ai}_{\scriptscriptstyle{\stackrel{78}{(\pm)}}} &=&0\,. 
%\end{eqnarray}
%

%The scalar fields $A^{Ai}_{\scriptscriptstyle{\stackrel{56}{(\pm)}}}$ are all massive fields 
%with the masses of the condensate scale (Table~\ref{Table con.}), while the scalar fields 
%$A^{Ai}_{\scriptscriptstyle{\stackrel{78}{(\pm)}}}$ change masses at the electroweak break.

Let me pay attention to the reader, that the term $\gamma^0\, \stackrel{78}{(-)} \, $
$ \tau^{Ai}$ $A^{Ai}_{\scriptscriptstyle{\stackrel{78}{(-)}}}$  in Eq.~(\ref{eigentau1tau2})
transforms the right handed $u_{R}^{c1}$ quark from the first line of Tables~\ref{Table so13+1.}%
--\ref{Table so13+1.b} 
into the left handed $u_{L}^{c1}$ quark from the seventh line of the same table~\footnote{This
transformation of the right handed family members into the corresponding left handed partners can 
easily be calculated by using Eqs.~(\ref{graphbinoms}, \ref{snmb:gammatildegamma}, 
\ref{plusminus}).}, which can, due to the properties of the scalar fields (Eq.~(\ref{checktau13Y})),
be interpreted also in the {\it standard model} way, namely, that
$A^{Ai}_{\scriptscriptstyle{\stackrel{78}{(-)}}}$ "dress" $u_{R}^{c1}$ giving it the weak and the
hyper charge of the left handed $u_{L}^{c1}$ quark, while $\gamma^0$ changes handedness. 
Equivalently happens to $\nu_{R}$ from the $25^{th}$ line, which transforms under the action of  
$\gamma^0\, \stackrel{78}{(-)}$ $ \tau^{Ai}$  $A^{Ai}_{\scriptscriptstyle{\stackrel{78}{(-)}}}$, 
into  $\nu_{L}$ from the $31^{th}$ line.
%which again can be interpreted in the {\it standard model} way: With the action of $\gamma^0$ 
%and by "dressing" $\nu_{R}$ at the same time by 
%$A^{Ai}_{\scriptscriptstyle{\stackrel{78}{(-)}}}$, transforming it into  $\nu_{L}$. 

The operator $\gamma^0\,\stackrel{78}{(+)}$ $\tau^{Ai}$ 
$ A^{Ai}_{\scriptscriptstyle{\stackrel{78}{(+)}}}$ transforms $d_{R}^{c1}$ from the third line of 
Tables~\ref{Table so13+1.}--\ref{Table so13+1.b}  into $d_{L}^{c1}$ from the  fifth line of this
 table, or $e_{R}$ 
from the $27^{th}$ line into $e_{L}$ from the $29^{th}$ line, where 
$ A^{Ai}_{\scriptscriptstyle{\stackrel{78}{(+)}}}$  belong to the scalar fields from
Eq.~(\ref{commonAi}).
%One can use in this two cases,  knowing the properties of the scalar fields
%(Eq.~(\ref{checktau13Y})), again the {\it standard model} interpretation, in which the scalar
% fields 
%$A^{Ai}_{\scriptscriptstyle{\stackrel{78}{(+)}}}$ "dress" $d_{R}^{c1}$ and $e_{R}$ with the
%weak and hyper charges of the left handed partners, while $\gamma^0 $ changes handedness. 

The term $\gamma^0\,\stackrel{78}{(\mp)}$ 
$\tau^{Ai}$ $A^{Ai}_{\scriptscriptstyle{\stackrel{78}{(\mp)}}}$ of the action
(Eqs.~(\ref{action}, \ref{eigentau1tau2})) determines the Yukawa couplings.
%All the scalar fields $A^{Ai}_{\scriptscriptstyle{\stackrel{78}{(-)}}}$, presented in
%Eq.~(\ref{commonAi}),  carry the weak and the hyper charge
% (Eqs.~(\ref{so42}, \ref{so64})) of the higgs of the {\it standard model}. 
The operator $\tau^{Ai}$, if representing the first three operators in
Eq.~(\ref{commonAi}),  (only) multiplies the right handed family member with its eigenvalue. 
If $\tau^{Ai}$ represents the last four operators of Eq.~(\ref{commonAi}), the operators 
$\gamma^0\,\stackrel{78}{(\mp)}$ $\tau^{Ai}$ 
$ A^{Ai}_{\scriptscriptstyle{\stackrel{78}{(\mp)}}}\;$ (($\mp$) for $(u_{R},\nu_{R})$ and 
$(d_{R},e_{R})$, respectively)    
transform the right handed family member of one family into the left handed partner of another 
family within the same group of four families, since these four operators manifest the symmetry
twice ($\widetilde{SU}(2)_{\widetilde{SO}(3,1)}\times \widetilde{SU}(2)_{\widetilde{SO}(4)}$). 
One group of four families carries  the family quantum numbers ($\vec{\tilde{\tau}}^{1}$, 
$\vec{\tilde{N}}_{L}$), the other group of four families carries the family quantum numbers
 ($\vec{\tilde{\tau}}^{2}$, $\vec{\tilde{N}}_{R}$). 

The nonzero vacuum expectation values of the scalar fields of Eq.~(\ref{commonAi}) break the 
mass protection mechanism of quarks and leptons and determine correspondingly the mass 
matrices (Eq.~(\ref{M0})) of the two groups of quarks and leptons.  

In loop corrections  all the scalar and vector gauge fields, which couple to fermions, contribute. 
Correspondingly all the off diagonal matrix elements of the mass matrix (Eq.~(\ref{M0})) 
depend on the family members quantum numbers.

That the scalar fields $A^{Ai}_{\scriptscriptstyle{\stackrel{78}{(\pm)}}}$  
are either {\it triplets} as the gauge fields of the {\it family quantum numbers} 
($\vec{\tilde{N}}_{R}, \,$ $\vec{\tilde{N}}_{L},\,$ $ \vec{\tilde{\tau}}^{2},\,$ 
$\vec{\tilde{\tau}}^{1}$);
% Eqs.~(\ref{so1+3tilde}, \ref{so42tilde}, \ref{bosonspin0})) 
or they are singlets as the gauge fields of 
$Q=\tau^{13}+Y, \,Q'= -\tan^{2}\vartheta_{1} Y$ $ + \tau^{13} $ and
 $Y' = -\tan^2 \vartheta_{2} \tau^{4} + \tau^{23}$, is shown in Ref.~\cite{JMP2015},
 Eq.~(22).

One finds 
\begin{eqnarray}
\label{checktildeNL3Q}
\tilde{N}_{L}^{3}\,\tilde{A}^{\tilde{N}_{L} \spm}_{\scriptscriptstyle{\stackrel{78}{(\pm)}}} &=&
\spm   \tilde{A}^{\tilde{N}_{L}\spm}_{\scriptscriptstyle{\stackrel{78}{(\pm)}}}\,,\quad
\tilde{N}_{L}^{3}\,\tilde{A}^{\tilde{N}_{L}3}_{\scriptscriptstyle{\stackrel{78}{(\pm)}}}=0\,,\nonumber\\
Q \,A^{Q}_{\scriptscriptstyle{\stackrel{78}{(\pm)}}} &=&0\,.
\end{eqnarray}
%
%One applies $\tilde{\tau}^{23}$,  $\tilde{\tau}^{13}$, $\tilde{N}_{R}^{3}$
%and $\tilde{N}_{L}^{3}$ on their eigenstates. 
%
with $ Q={\cal S}^{56} + {\cal \tau}^{4}= {\cal S}^{56} -\frac{1}{3}({\cal S}^{9\,10}+
{\cal S}^{11\,12} + {\cal S}^{13\,14})$, and with ${\cal \tau}^{4}$ defined in the footnote
on the page of Eq.~(\ref{tau}), if replacing $S^{ab}$ by ${\cal S}^{ab}$ from 
Eq.~(\ref{bosonspin0}). 

Similarly one finds properties with respect to the $Ai$ quantum numbers for all the scalar fields
$A^{Ai}_{\scriptscriptstyle{\stackrel{78}{(\pm)}}}$. 

All  other scalar fields: $A^{Ai}_{s}\,, s\in(5,6)$ and  $A^{Ai}_{t t'}\,, (t,t') \in(9, \dots,14)$
have masses of the order of the condensate scale and contribute to matter-antimatter 
asymmetry~\cite{norma2014MatterAntimatter}. 

\section{Mass matrices and properties of quarks~\cite{gmdn07,gn2013,gn2015}}
\label{massmatrices}

One of the most important open questions in the elementary particle physics is: Where do 
the families originate? Explaining the origin of families would give us the answer about the 
number of families which will be observed at the low energy regime, about the origin 
of the scalar field(s) and the Yukawa couplings, telling how many scalar fields can we 
expect to observe at the acceptable energies and would also explain differences in 
the fermions properties - the differences in masses and mixing matrices among family members --
quarks and leptons, as well as it might explain the hierarchy problem.

The {\it spin-charge-family} theory predicts that there are at the low energy regime twice 
four families of quarks and leptons. This means that besides the observed three there is the 
fourh family of quarks and leptons. It also mean that the stable of the upper four families 
must also be observed. I shall comment the agreement of the existence of the 
fourth family quarks with the observations, in particular with the contribution of the fourth family
to the production of the higgs in the quark-fusion process, in Sect.~\ref{fourthfamily}.
The properties of the lowest of the upper four families, contributing to the dark matter, will be 
shortly presented in Sect.~\ref{dark matter} \cite{gn}. 

The mass matrix of any of the family member (quark or lepton)  demonstrates in the 
massless basis the $U(1)\times \widetilde{SU}(2)\times$ $\widetilde{SU}(2)$ symmetry
(each of the two $\widetilde{SU}(2)$ is a subgroup, one of $\widetilde{SO}(3,1)$ and the 
other of $\widetilde{SO}(4)$) symmetry),  Eq.~(\ref{M0}). 

To the masses of the lower and upper four families all the scalars with the family members 
quantum numbers $(Q,Q',Y')$ contribute, while the scalars with the family quantum numbers 
split the eight families into twice four families: To the masses of the lower four families the scalar 
fields, which are the gauge fields of $\vec{\tilde{N}}_{L}$ and  $\vec{\tilde{\tau}}^{1}$ 
contribute. To the masses of the upper four families the gauge fields of $\vec{\tilde{N}}_{R}$ 
and  $\vec{\tilde{\tau}}^{2}$ contribute. 
\begin{small}
 \begin{equation}
 \label{M0}
 {\cal M}^{\alpha} = \begin{pmatrix} 
 - a_1 - a & e     &   d & b\\ 
 e     & - a_2 - a &   b & d\\ 
 d     & b     & a_2 - a & e\\
 b     &  d    & e   & a_1 - a
 \end{pmatrix}^{\alpha}\,.
 \end{equation}
 \end{small}

Although any accurate $3\times 3$ submatrix of the $4 \times 4$ unitary matrix determines the 
$4 \times 4$ matrix uniquely, neither the quark  nor (in particular) the lepton 
$3\times 3$ mixing matrix are measured accurately enough that it would be possible to determine 
three complex phases of the $4 \times 4$ mixing matrix as well as the mixing matrix elements 
of the fourth family members to the lower three.  We therefore assumed in our 
calculations~\cite{gmdn07,gn2013,gn2015} that the mass matrices are symmetric 
and real. Correspondingly the mixing matrices are orthogonal. We fitted the $6$ free parameters 
of each quark mass matrix, Eq.~(\ref{M0}),  to twice three measured quark masses ($6$), 
and to the $6$ (from the experimental data extracted) parameters of the 
corresponding $4 \times 4$ mixing matrix.

While the experimental accuracy of the quark masses of the lower three families does not influence 
the calculated mass matrices considerably, it turned out that the experimental accuracy of the 
$3\times 3$ quark mixing matrix is not good enough to trustworthy determine the mass intervals 
for the fourth family quarks.

Taking into account our calculations fitting
the experimental data (and the meson decays evaluations in literature, as well as our own)
we estimated that the fourth family quarks masses might be above $1$ TeV. 
Choosing the masses of the fourth family quarks we were able  not only to calculate the fourth
family matrix elements to the lower three families, but also predict towards which values will the
matrix elements of the $3\times 3$  submatrix move in the more accurate
experiments~\cite{gn2015}.

The higher are the fourth family quark masses the closer are the mass matrices to the democratic 
ones. The fourth family quark masses are closer to each other, the smaller is the contribution
of the scalar fields with the family members quantum numbers to the fourth family masses.

The complex mass matrices would lead to unitary and not to orthogonal mixing matrices.
The more accurate experimental data for quarks mixing matrix would allow us to extract also 
the phases of the unitary mixing matrix, allowing us to predict the fourth family masses.

\section{Triplets with respect to space index $s=(9,\dots,14)$ and the matter-antimatter 
asymmetry in the universe}
\label{triplets}

The gauge fields with the space index $t \in (9,\dots,14)$ form the triplets and antitriplets with
 respect to the space index $s=(9,\dots,14)$ (one triplet and one antitriplet for each $Ai$). 
I kindly ask the reader to study this topic  in Ref.~\cite{norma2014MatterAntimatter}.

The colour triplet scalars namely contribute to transition from antileptons into quarks and 
antiquarks into quarks and back, unless the scalar condensate of the two right handed neutrinos, 
presented in Table~\ref{Table con.}, breaks matter-antimatter 
symmetry~\cite{norma2014MatterAntimatter}, 
offering the explanation for the matter-antimatter asymmetry in our universe.  The colour triplet 
scalars and the condensate cause also the decay of proton~\cite{norma2014MatterAntimatter}.

The condensate breaks also the $SU(2)_{II}$ symmetry, 
leaving massless besides gravity only the colour, weak and the hyper charge vector gauge fields,
the corresponding charges remain conserved. 
Also all scalar fields get masses through the interaction with the condensate.

There are no additional scalar indices and therefore no additional corresponding scalars with respect 
to the scalar indices in this theory. 
Scalars, which do not get nonzero vacuum expectation values, keep masses on the condensate 
scale.

\section{The fourth family quarks and their couplings to the scalar fields}
\label{fourthfamily}

The {\it spin-charge-family} theory predicts the fourth family to the observed three, while there
has been no direct observation of the fourth family quarks with the masses below $1$ TeV. The 
fourth family quarks with masses above $1$ TeV contribute according to the {\it standard model} 
(the {\it standard model} Yukawa couplings of quarks to the scalar higgs is proportional to  
$\frac{m^{\alpha}_{4}}{v}$, where $m^{\alpha}_{4}$ is the fourth family member 
($\alpha=u,d$) mass and $v$ the vacuum expectation value of the scalar) to either the 
quark-gluon fusion production of the scalar field (the higgs) or to the scalar field decay into two 
photons $\approx 10$ times too much in comparison with the observations. Correspondingly 
the high energy physicists do not expect the existence of the fourth family members 
at all~\cite{AhmedAliMatthiasNeubert}.

I am stressing~\cite{norma2016higgs} in this section that the $u_i$-quarks and $d_i$-quarks 
of all the families if coupled with the opposite sign to the scalar fields, carrying the family 
quantum numbers,  $\tilde{A}^{\tilde{A}i}_{\pm} (\tilde{Ai}=(\tilde{\tau}^{1i}, \tilde{N}^{i}_{L}$), 
(Eq.~\ref{commonAi})) (they are the same for all the family members) do not contribute to either 
the quark-gluon fusion production of the scalar fields with the family quantum numbers or to the 
decay of 
these scalars into two photons, if the $u_i$-quarks and $d_i$-quarks have the same mass. 
Since the $u_4$-quarks and $d_4$-quarks might have similar masses, if their masses are mostly 
determined by the scalars with the family quantum numbers, the observations are consequently 
not in 
contradiction with the {\it spin-charge-family} theory prediction that there exists the fourth family
coupled to the observed three.

The  couplings of $u_i$ and $d_i$ to the scalars carrying the family members quantum numbers 
are determined besides by the corresponding couplings by also by theeigenvalues of the operators 
$(Q,Q',Y')$ on the quarks states. 
%The fact that the masses of the observed three families depend strongly on the family members 
%quantum numbers means that their masses 
%
The strong influence of the scalar fields carrying the family members quantum numbers on the 
masses of the lower (observed) three families manifests in the huge differences in the  masses 
of $u_i$ and $d_i$, $i=(1,2,3)$, among families ($i$) and family members ($u,d$). For the fourth
family quarks, which are more and more decoupled from the observed three families the higher 
are their masses~\cite{gn2015,gn2013}, the influence of the scalar fields carrying the family 
members quantum numbers on their masses is expected to be much weaker. 
Correspondingly the $u_4$ and $d_4$ masses become closer to each other the higher are their 
masses and the weaker is their couplings (the mixing matrix elements) to the lower three families.

If the masses of the fourth family quarks are close to each other, then $u_{4}$ and 
$d_{4}$ contribute in the quark-gluon fusion  very little to the production of the scalar field - the 
higgs - which is mostly  superposition of the scalar fields with the family members quantum numbers,
what is in agreement with the observation: In the quark-gluon fusion production of the higgs 
mostly the top ($u_3$) contributes.

 The $u_{4}$ and $d_{4}$ quarks of almost the same mass couple weakly to scalars which carry 
the family members quantum numbers and correspondingly contribute weakly to  the production 
%of higgs - the scalar 
%which is mostly a superposition of scalars with the family members quantum numbers. 
%. 
%%, which is in the {\it spin-charge-family} theory obviously mostly the superposition of the scalar 
% fields with the family members quantum numbers,

The $u_{4}$ and $d_{4}$  can still, but only weakly, contribute to the production of
some of the remaining superposition of the scalar fields, predicted by the {\it spin-charge-family}
theory, most probably  to one of those, which is mainly the superposition of scalar fields carrying the 
family quantum numbers. This might be the scalar with the mass of $\approx 750$ GeV, observed 
at the  LHC~\cite{twophotonsCMS,twophotonsATLAS}, if this is a real event.

Let me write the statement~\cite{norma2016higgs}.

\noindent
{\it Statement 2:} The $u_i$-quarks and $d_i$-quarks of equal masses, which couple with the 
opposite sign to the scalar fields, carrying the family quantum numbers 
($\tilde{A}^{\tilde{A}i}_{\pm}, \vec{\tilde{A}}=(\vec{\tilde{\tau}}^{1}, \vec{\tilde{N}}_{L})$, 
(Eq.~\ref{commonAi})), can not contribute in the quark-gluon fusion to the production of these 
scalar fields.

%All $u_i$  couple to  $d_i$ with the opposite sign to the scalar fields carrying the family quantum 
%numbers,
%% but also to all scalars with $s=(7,8)$, 
%but while the family quantum numbers do not 
%distinguish among the family members, the family members quantum numbers ($Q,Q',Y'$) do,
%what influences the coupling constants of quarks and leptons to scalar fields carrying the family 
%members quantum numbers in addition.
 
Let me briefly comment {\it Statement 2},  making use of  the  
technique~\cite{hn02,hn03}, discussed in App.~\ref{technique}. 
%, to present fermions in an transparent way. 

%
From Tables~\ref{Table so13+1.}--\ref{Table so13+1.b} we can see  presentation of spinors in
 terms of projectors and 
nilpotents, and read their properties. We need here their weak and hyper charges, while
taking into account that antiquarks, Tables~\ref{Table so13+1.}--\ref{Table so13+1.b}, carry the 
opposite charges than 
the corresponding quarks and that one can obtain the antiquarks as well by the application of the
operator~\cite{discretesym}
%
%\begin{eqnarray} 
%\label{CP}
$\mathbb{C}_{{\cal N}}  {\cal P}_{{\cal N}} =\gamma^0
 \prod_{\Im \gamma^s, s=5}^{d} \gamma^s\,\,
I_{\vec{x}_3} \,I_{x^6,x^8,\dots,x^{d}}$ on the corresponding quarks\footnote{
%\end{eqnarray}
%
Here $\gamma^0$ and $\gamma^1$ are real, $\gamma^2$ imaginary, 
$\gamma^3$ real, $\gamma^5$ imaginary, $\gamma^6$ real, alternating imaginary and real 
up to $\gamma^d$, which is in even dimensional spaces real.  $\gamma^a$'s appear in the 
ascending order.
Operators $I$ operate as follows: $\quad I_{x^0} x^0 = -x^0\,$; $
I_{x} x^a =- x^a\,$; $  I_{x^0} x^a = (-x^0,\vec{x})\,$; $ I_{\vec{x}} \vec{x} = -\vec{x}
\,$; $I_{\vec{x}_{3}} x^a = (x^0, -x^1,-x^2,-x^3,x^5, x^6,\dots, x^d)\,$; 
$I_{x^5,x^7,\dots,x^{d-1}}$ $(x^0,x^1,x^2,x^3,x^5,x^6,x^7,x^8,
\dots,x^{d-1},x^d)$ $=(x^0,x^1,x^2,x^3,-x^5,x^6,-x^7,\dots,-x^{d-1},x^d)$;
 $I_{x^6,x^8,\dots,x^d}$ 
$(x^0,x^1,x^2,x^3,x^5,x^6,x^7,x^8,\dots,$ $x^{d-1},x^d)$
$=(x^0,x^1,x^2,x^3,x^5,-x^6,x^7,-x^8,\dots$ $,x^{d-1},-x^d)$, $d=2n$.}. 
In Tables~\ref{Table so13+1.}--\ref{Table so13+1.b}  the phases of all the states are chosen to be
1. In this talk I use different phases, those presented in footnote ${}^i$, which enable the usual
presentation of fermions under the change of spin and under 
$\mathbb{C}_{{\cal N}}$ $\cdot {\cal P}_{{\cal N}}$.

\begin{figure}
\centering
\subfigure[]{
\includegraphics{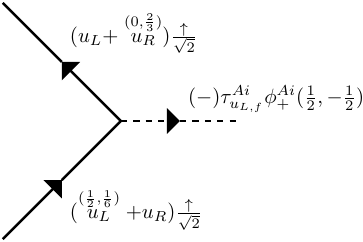}} \quad
\subfigure[]{
\includegraphics{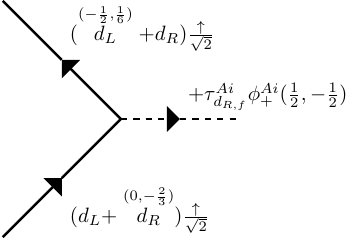}}\\
\vspace{4mm}
\subfigure[]{\includegraphics{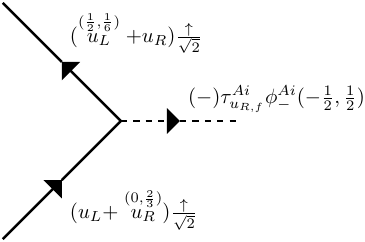}}\quad
\subfigure[]{\includegraphics{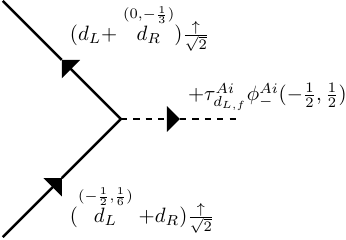}}
\caption{\label{Fig.1} %
The contributions of $u$ and $d$ quarks to the production of the scalar
fields $\Phi^{Ai}_{-}$ and  $\Phi^{Ai}_{+}$, when $\tau^{Ai}$ represent the family 
quantum numbers (which are for the lower four families $\tilde{\tau}^{1i}$ and 
$\tilde{N}^{i}_{L}$) or the family members quantum numbers ($Q, Q', Y'$), are presented:
 (a) the $u$-quark contribution to the scalar fields $\Phi^{Ai}_{+}$,
 (b) the   $d$-quark contribution to $\Phi^{Ai}_{+}$,
 (c) the  $u$-quark contribution to $\Phi^{Ai}_{-}$,
 (d) the  $d$-quark contribution to  $\Phi^{Ai}_{-}$. 
%%new
$\tau^{Ai}_{u_(L,R),f}$ and $\tau^{Ai}_{d_(L,R),f}$ denote the application values of the 
operators $\tilde{\tau}^{1i}$ and $\tilde{N}^{i}_{L}$ and  $Q, Q', Y'$ on the states. While 
$\tilde{\tau}^{1i}$ and $\tilde{N}^{i}_{L}$ do not distinguish among family members $u$ and
$d$ so that in this case the contribution of $u$ and $d$ have opposite signs, $Q, Q', Y'$ do,
 influencing the signs in addition. 
%%
%The index $Ai$ is not shown. It is not difficult to see from the figures that $u$-quarks and
%$d$-quarks contribute with opposite signs: Sign($g^u$)= - Sign($g^d$), 
%for any family to either
% $\Phi^{Ai}_{-}$ or to  $\Phi^{Ai}_{+}$ , $g^u$ and $g^d$ are the coupling constants  of 
%the $u$ and $d$ quarks to the scalar fields with the family quantum numbers, respectively. 
%The same is valid also for the scalar fields carrying the family members quantum numbers,
%but there also the eigenvalues of $(Q,Q',Y')$ on $u$ and $d$ must be taken into account,
%while the family quantum numbers are the same for $u$ and $d$. 
}
\end{figure}

In Table~\ref{udprop.}, the properties of $u$ and $d$
(and their antiquarks) needed in  Fig.~\ref{Fig.1}, are presented. 
\begin{table}
\tbl{%\label{udprop.} 
The weak, hyper and elm charges for quarks and antiquarks in their massless basis are presented, 
the colour charge is not shown. These and other properties of quarks 
and antiquarks, leptons and antileptons can be read from 
Tables~\ref{Table so13+1.}--\ref{Table so13+1.b} } 
% \begin{center}
 %\begin{small}
{ \begin{tabular}{c| r r r |c| r r r}
 \hline
 state & $ {\cal \tau}^{13}$ & $Y$ & $Q$ & antistate& $ {\cal \tau}^{13}$ &  $Y$ & $Q$\\
 \hline
% $A^{Ai}_{\scriptscriptstyle{\stackrel{78}{(-)}}}$ & $A^{Ai}_{7}+iA^{Ai}_{8}$& $+\frac{1}{2}$& 
 $        u_{Ri}$ & $ 0 $              & $\frac{2}{3}$  & $\frac{2}{3}$ &
 $\bar{u}_{Ri}$ & $-\frac{1}{2}$ & $-\frac{1}{6}$ & $-\frac{2}{3}$ \\
 $        u_{Li}$  & $ \frac{1}{2}$ & $ \frac{1}{6}$  & $\frac{2}{3}$ &
 $\bar{u}_{Li}$  & $ 0 $             & $-\frac{2}{3}$ & $-\frac{2}{3}$ \\
 $        d_{Ri}$  & $ 0 $             & $-\frac{1}{3}$  & $-\frac{1}{3}$ &
 $\bar{d}_{Ri}$ & $ \frac{1}{2}$ & $-\frac{1}{6}$ & $ \frac{1}{3}$ \\
 $        d_{Li}$  & $-\frac{1}{2}$ & $ \frac{1}{6}$  & $-\frac{1}{3}$ &
 $\bar{d}_{Li}$  & $ 0 $             & $ \frac{1}{3}$ & $\frac{1}{3}$\\
\hline
\end{tabular}
}
 %\end{small}
% \end{center}
\label{udprop.}
 \end{table}

In Fig.~\ref{Fig.1} the properties of the $u$ and $d$ quarks, contributing to 
the production of the dynamical part of the scalar fields -
$\Phi^{Ai}_{\stackrel{78}{(\pm)}}$,~Eq.(\ref{commonAi}), ($A^{Ai}_{\stackrel{78}{(\pm)}}$
$=$ $\Phi^{Ai}_{\stackrel{78}{(\pm)}}$ +$v^{Ai}_{\stackrel{78}{(\pm)}}$ ) - 
in the quark-gluon fusion, are presented.  
% $\Phi^{Ai}_{\stackrel{78}{(\pm)}}$ is presented
%only as $\Phi_{+}$ and $\Phi_{-}$ to make the figure more readable.
%
One notices the opposite signs of the couplings of $u_{i}$ with respect to $d_{i}$ for ether
$\Phi_{-}$ or for $\Phi_{+}$. 
%This is the case for either the two triplets  with the family quantum
%numbers as well as for the three singlets with the family members quantum numbers.
%

Let us  discuss the possibility that at the LHC~\cite{twophotonsCMS,twophotonsATLAS} 
observed two photons effect, if trustworthy, might have the origin in one 
of the nine mass eigenstates of the scalar fields, contributing to mass matrices~(Eq. (\ref{M0})) 
of the lower group of four families. 
Namely, if the masses of $u_{4}$ and $d_{4}$ are not completely equal (while the couplings 
of these four family quarks contribute with the opposite signs to the production of scalars with 
the family quantum numbers as presented in Fig.~\ref{Fig.1}), then besides the 
observed scalar field (the higgs) also another mass eigenstate, the superposition of mostly
the scalar fields with the family quantum numbers, can be  produced. 
Such a scalar decays mostly through the production of two photons, since the mixing matrix 
elements of the fourth family quarks to the lower (observed) three families are very 
small~\cite{gn2013,gn2015}. If the event, reported in 
Refs.~\cite{twophotonsCMS,twophotonsATLAS}, is a true one, this scalar field would have
a mass of $\approx$ $750$ GeV.

\begin{figure}
\centering
\subfigure[]{
\includegraphics{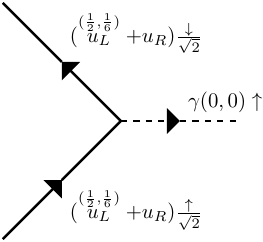}} \qquad
\subfigure[]{
\includegraphics{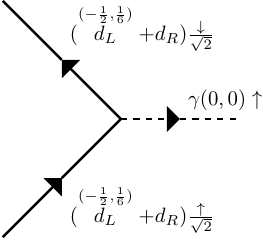}}
\caption{\label{Fig.2Gamma} %
The contribution of quarks to the production of   a photon is presented, to manifest the difference
in the production of the scalar fields and photons.}
%\label{Fig.2Gamma} 
\end{figure}

The production of a photon by quarks is presented in 
Figs.~\ref{Fig.2Gamma} (a, b) for $u_{L}$ and $d_{L}$ with spin up, as an example. 
Only the weak ($\tau^{13}$) and hyper ($Y$) charges besides the spin ($S^{12}$, presented 
in figures by arrows) are shown.  In this case the phases are, as expected, the same for 
$u$ and $d$, up to the sign determined by the electromagnetic charge, which are for $u$ 
and $d$ different.  
Equivalent figures can be drawn for gluons, where the colour charge replaces the electromagnetic
charge of quarks~\cite{norma2016higgs}.
The figures are valid for any $Ai$ and correspondingly also for any superposition of 
$\Phi^{Ai}_{\pm}$.

\section{The upper groups of four families and the dark matter}
\label{dark matter}

As discussed in Sect.~\ref{SCFT} the {\it spin-charge-family} theory~\cite{NBled2013,NBled2012,%
norma92,norma93,norma94,pikanorma,portoroz03,JMP,norma95,gmdn07,gn,gn2013,gn2015,NPLB,%
N2014scalarprop,norma2014MatterAntimatter,JMP2015} predicts in the low energy region two 
decoupled groups of four families. In Ref.~\cite{gn} the possibility that the dark matter consists 
of clusters of the fifth family - the stable heavy family of quarks and leptons (with zero Yukawa
couplings to the lower group of four families) - is discussed.

We made in Ref.~\cite{gn} a rough estimation of the properties of  baryons  of this fifth family 
members, of their behaviour during the evolution of the universe and when scattering on 
the ordinary matter. We studied possible limitations on the family properties due to the 
cosmological evidences, the direct experimental evidences, and all others (at that time) known 
properties of the dark matter.

We used the simple hydrogen-like model to evaluate the properties of these heavy baryons and 
their interaction  among themselves and with the ordinary  nuclei, taking into account that for 
masses of the order of $1$ TeV or larger the one gluon exchange determines the force among
the constituents of the fifth family baryons. Due to their very large masses "the nuclear interaction"
among these baryons has very interesting properties.
We concluded that it is the fifth family neutron, which is very probably the most stable nucleon.

We followed the behaviour of the fifth family quarks  and antiquarks in the plasma of the 
expanding universe, through the freezing out procedure, through the colour phase transition, while 
forming neutrons, up to the present dark matter.  Also the scattering of the fifth family 
neutrons among themselves and on the ordinary matter was evaluated.

The cosmological evolution suggested the limits for the masses of the fifth family quarks
\begin{eqnarray}
\label{massrange}
 10 \; {\rm TeV}  < m_{q_5} \, c^2 < {\rm a \, few} \cdot 10^2\, {\rm TeV} 
\end{eqnarray}
and for the  scattering cross sections
\begin{eqnarray}
\label{scatteringrange} 
 10^{-8}\, {\rm fm}^2\, < \sigma_{c_5}\, <   10^{-6} \,{\rm fm}^2 \,, 
\end{eqnarray} 
while the measured density of  the  dark matter does not put much limitation on the properties of 
heavy enough clusters.  

The direct measurements limited the fifth family quark mass to~\cite{gn}
\begin{eqnarray}
\label{direct}  {\rm several}\, 10 \,{\rm TeV} < m_{q_{5}}c^2 < 10^5\, {\rm TeV}\,. 
\end{eqnarray} 

We also find that our fifth family baryons of the mass of a few hundreds TeV/${c^2} $ 
have  for a factor more than $100$ times too small scattering amplitude with the ordinary matter
to cause a measurable heat flux on the Earth's surface.

\section{Concluding remarks}
\label{conclusion}

One of the most important open questions in the elementary particle physics is: {\it Where do 
the families originate?} Explaining the origin of families would answer the question about the 
number of families which are possibly observable at the low energy regime, about the origin 
of the scalar field(s) and the Yukawa couplings (the couplings of fermions to the scalar field(s)), 
about the differences in the fermions properties - the differences in the masses and mixing 
matrices among family members --
quarks and leptons, as well as about the hierarchy in quark and lepton masses.  

To understand the history of the universe is needed to explain the assumptions of the 
{\it standard model}, as well as the phenomena like the existence of the dark matter, dark 
energy and matter/antimatter asymmetry.

I demonstrated in this talk, that the {\it spin-charge-family} theory, starting with the simple action
in $d=(13+1)$ for fermions which carry two kinds of spins (no charges) and couple correspondingly 
to the vielbeins and the two kinds of spin connection fields and the corresponding boson 
fields~\footnote{If there are no spinors present, are the two spin connections expressible 
uniquely with the vielbeins~\cite{DNproof}.}, offers the explanation for all the assumptions of the
 {\it standard model}:\\
{\bf a.} $\;\,$ The theory explains all the properties of the family members - quarks and leptons,
left and right handed, and their right and left handed antiquarks and antileptons~\footnote{One
Weyl representation of $SO(13+1)$ contains, if analyzed with respect to the {\it standard model }
groups, all the members of one family, the coloured quarks and colourless leptons, and the 
anticoloured antiquarks and (anti)colourless antileptons, with the left handed leptons  
carrying the weak charge and the right handed ones weak chargeless, while the left handed 
antispinors are weak chargeless and the right handed ones carry the weak charge.}, explaining why
 the left handed spinors carry the weak charge while the right handed do not (the right handed 
neutrino is the regular member of each family).\\ 
{\bf b.} $\;\,$  It explains the appearance and the properties of the families of family members. \\
{\bf c.} $\;\,$ It explains the existence of the gauge vector fields of the family members charges. \\
{\bf d.} $\;\,$ It explains the appearance and the properties of the scalar field (the higgs) and the 
Yukawa couplings.\\

\noindent
It also offers the explanation for the phenomena, which are not integrated into the 
{\it standard model}, like:\\ 
{\bf e.} $\;\,$ It explains the existence of the dark matter.\\
{\bf f.} $\;\,$ It explain the origin of the (ordinary) matter/antimatter asymmetry in the universe. 

\noindent
The theory predicts:\\
{\bf g.} $\;\,$ There are twice two groups of four families of quarks and leptons at low energies.\\  
{\bf g.i.} The fourth family  with masses above $1$ TeV, weakly coupled to the observed
three families, will be measured at the LHC.\\ 
{\bf g.ii.} The quarks and leptons of the fifth family - that is of the stable one of the upper four 
families - form the dark matter. The  family members, which form the chargeless clusters, manifest,
 due to their very heavy masses, a "new nuclear force".\\
{\bf h.} $\;\,$ The predicted scalar fields with the space index $(7,8)$ are doublets with respect to
the space index (carrying the weak and the hyper charge of the {\it standard model} higgs).
They carry in addition:\\ 
 {\bf h.i.} Either they carry one of the three family members quantum numbers, 
$(Q,Q',Y')$ - belonging correspondingly  to one of three singlets. \\
 {\bf h.ii.} Or they carry family quantum numbers - belonging correspondingly  to one of the
 twice two triplets.\\
{\bf h.iii.} The  three singlets and the two triplets determine mass matrices of the lower 
four families, contributing to masses of the heavy vector bosons. \\ 
{\bf h.iv.} These scalars determine the observed higgs and the Yukawa couplings.\\
{\bf i.} $\;\,$ The predicted scalar fields with the space index $(9,10,..,14)$ are triplets with respect to 
the space index. They cause the transitions from antileptons into quarks and antiquarks into quarks 
and back. The condensate breaks the matter/antimatter symmetry, causing the asymmetry in the 
(ordinary) matter with respect to antimatter. \\ 
{\bf i.i.} The condensate is responsible also for the proton decay.\\
{\bf j.} $\;\,$ The condensate is a scalar of the two right handed neutrinos with the family quantum 
numbers of the upper four families. \\
{\bf k.} $\;\,$ The condensate gives masses to all the gauge fields with which it 
interacts.\\
{\bf k.i.}  It gives masses to all scalar fields and to vector fields, leaving massles only the colour, 
the weak, the hyper vector gauge fields and the gravity in ($(3+1)$). \\
{\bf l.} $\;\,$ There is the $SU(2)$ (belonging together with the weak $SU(2)$ to $SO(4)$ 
gauge  fields included in $SO(7,1)$) vector gauge field, which gain masses of the order of the 
appearance of the condensate.\\ 
{\bf m.} $\;\,$ At the electroweak break the scalar fields with the space index $(7,8)$ change their
 mutual interaction, and gaining nonzero vacuum expectation values, break the weak and the 
hyper charges and correspondingly the mass protection of fermions, making them massive.\\
{\bf n.} $\;\,$ The symmetry of mass matrices allow, in the case that the experimental data 
for the mixing submatrix $3 \times 3$ of the $4 \times 4$ mixing matrix are accurately enough, to 
determine the mixing matrix and the masses of the fourth family quarks. The accuracy, with 
which the masses of the six lower families are measured so far, does not influence the results
appreciably. Due to uncertainty of the experimental data for the $3\times 3$ mixing submatrix 
we are only able to determine the $4 \times 4$ quark mixing matrix for a chosen masses of the
the fourth family quarks. However, we also predict how will the $3 \times 3$ submatrix of the 
mixing matrix change with more accurate measurements.\\
{\bf n.i.} The fourth family quarks mass matrices are for masses above $1$ TeV closer and 
closer to the democratic matrices. The less the scalars with the family members quantum
numbers contribute to masses of the fourth family quarks, the closer is $m_{u_4}$ to
 $m_{d_4}$.\\
{\bf n.ii.} The large contribution of the scalars with the family members quantum numbers 
($Q,Q',Y'$) to the masses of the lower four families manifests in the large differences 
of quarks masses of the lower four families.\\
{\bf n.iii.} Although we have done calculations also for leptons, must further analyses of their 
properties wait for more accurate experimental data.\\
{\bf o.} $\;\,$ In the case that  the $u_{4}$ and $d_{4}$ quarks have similar masses -
determined mostly by the scalar fields carrying the family quantum numbers - they contribute
mostly to the production of these scalars, while their contribution to the 
production of those scalars which carry the family members quantum numbers - to the higgs in 
particular - is much weaker, which is in 
agreement with the experiment~\footnote{The 
coupling constants of the singlet scalar fields differ among themselves and also from the coupling 
constants of the two triplet scalar fields}.\\
%{\bf o.i.} Since the scalar fields carrying the family members quantum numbers influence obviously  
%the masses of the lower three families strongly,  the fourth family quarks
%%, the masses of which are influenced mostly by the scalar fields with the family quantum numbers, have almost equal 
%masses, contributing  correspondingly little
%to the production of the observed higgs, in agreement with the observations. \\
{\bf o.i.} The fourth family quarks might contribute to the production of the scalar field, which  
is a superposition of mostly the scalars carrying the family quantum numbers. The event,
observed at the LHC as the two photon production, might be the signal of this new scalar
field with the mass of $\approx 750$ GeV.\\
{\bf p.} $\;\,$ All the degrees of freedom discussed in this talk are already a part of the simple 
starting action~Eq.(\ref{action}).\\
{\bf p.i.} The way of breaking symmetries (ordered by the conditions determining the history 
of our universe) is assumed so that it leads in $d=(3+1)$ to the observable symmetries.\\
{\bf p.ii.} Also the effective interaction among scalar fields is assumed, although we could 
derive it in principle from the starting action.\\

\noindent
There are several open problem in the {\it spin-charge-family} theory:\\
{\bf r.} $\;\,$ Since this theory is, except that fermions carry two kinds of spins - one
kind taking  care of spin and charges, the second one taking care of families - a kind of 
the Kaluza-Klein theories, it shares at very high energy with these theories the quantization
problem. \\  
{\bf s.} $\;\,$ The dimension of space-time, $d=(13+1)$, is in the {\it spin-charge-family}
theory chosen, since $SO(13,1)$ contains all the members, assumed in the {\it standard 
model}. It contains also the right handed neutrino (which carries the $Y'$ quantum number.)\\
{\bf s.i.} It must be shown, however, how has nature "made the decision" in evolution 
to go through this dimension and what is indeed the dimension of space-time (infinite?).\\
{\bf t.} $\;\,$ There are many other open question, like: \\
{\bf t.i.}  What is the reason for the (so small) dark energy?\\ 
{\bf t.ii.} At what energy the electroweak phase transition occurs?\\
{\bf t.iii.} Why do we have fermions and bosons?\cite{hnfermionization2015}\\
{\bf t.iv.} Some of them might be solved by comparison with other approaches and theories. 

% See the talk. fermionization

%{quark-gluon fusion and the decay of the scalar coupled to the fourth family quarks to two 
%photons}
%Couplings o an additional $SU(2)$ triplet vector gauge fields (belonging together with the 
%f quarks to the scalar fields }
%\label{gluonquarkscalarquarktwophotons} 
%\section{Make a choice}

% \section*{Acknowledgements}.

%\begin{appendix} [Optional Appendix Title]

%Correct the references
\begin{appendix}[]
\section{Short presentation of spinor technique~\cite{JMP2015,norma93,hn02,hn03}}
\label{technique}

This appendix is a short review (taken from~\cite{JMP}) of the technique~\cite{norma93,DKhn,%
hn02,hn03}, 
initiated and developed in Ref.~\cite{norma93}, while  proposing the {\it spin-charge-family} 
theory~\cite{NBled2013,NBled2012,norma92,norma93,norma94,%
pikanorma,portoroz03,JMP,norma95,gmdn07,gn,gn2013,gn2015,NPLB,N2014scalarprop,%
norma2014MatterAntimatter,JMP2015}.
All the internal degrees of freedom of spinors, with 
family quantum numbers included, are describable in the space of $d$-anticommuting (Grassmann) 
coordinates~\cite{norma93}, if the dimension of ordinary space is also $d$. % (App.~\ref{lorentz}). 
There are two kinds of operators in the Grassmann space fulfilling the Clifford algebra and 
anticommuting with one another~Eq.(\ref{gammatildegamma}). The technique  was further 
developed in the present 
shape together with H.B. Nielsen~\cite{DKhn,hn02,hn03}. 

In this last stage we  rewrite a spinor basis, written in Ref.~\cite{norma93} as products of polynomials 
of Grassmann coordinates of odd and even Grassmann character, chosen to be eigenstates of the Cartan 
subalgebra defined by the two kinds of the Clifford algebra objects, as products of nilpotents and 
projections, formed as odd and even objects of $\gamma^a$'s, respectively, and  chosen to be eigenstates 
of a Cartan subalgebra of the Lorentz groups defined by $\gamma^a$'s and $\tilde{\gamma}^a$'s.   

The technique can be used to construct a spinor basis for any dimension $d$
and any signature in an easy and transparent way. Equipped with the graphic presentation of basic states,  
the technique offers an elegant way to see all the quantum numbers of states with respect to the two 
Lorentz groups, as well as transformation properties of the states under any Clifford algebra object. 

Ref.~\cite{JMP2015}, App.~B, %\ref{lorentzappendix} 
briefly represents the starting point~\cite{norma93} of this technique. 
%in order to better understand the Lorentz transformation properties of both
There are two kinds of the Clifford algebra objects, 
$\gamma^a$'s and $\tilde{\gamma}^a$'s.%, as well as of spinor, vector, tensor and scalar fields, 
%appearing in the {\it spin-charge-family} theory, that is of the vielbeins and spin connections of 
%both kinds, $\omega_{ab \alpha}$ and $\tilde{\omega}_{ab\alpha}$, and of spinor fields, family members and
%families.

These objects %$\gamma^a$ and $\tilde{\gamma}^a$ 
have properties,
\begin{eqnarray}
\label{gammatildegamma}
&& \{ \gamma^a, \gamma^b\}_{+} = 2\eta^{ab}\,, \quad\quad    
\{ \tilde{\gamma}^a, \tilde{\gamma}^b\}_{+}= 2\eta^{ab}\,, \quad,\quad
\{ \gamma^a, \tilde{\gamma}^b\}_{+} = 0\,.
\end{eqnarray}

If $B$ is a Clifford algebra object, let say a polynomial of $\gamma^a$,
%(Eq.~(\ref{gammavector})),  
$B=a_{0} + a_{a}\,\gamma^a + a_{a b}\, \gamma^a \gamma^b + \cdots + 
a_{a_1 a_2 \dots a_d}\, 
\gamma^{a_1}\gamma^{a_2}\dots \gamma^{a_d}$, 
one finds 
\begin{eqnarray}
(\tilde{\gamma}^a B : &=& i(-)^{n_B} \,B \gamma^a \,) \;|\psi_0 \,>, \nonumber\\
B &=& a_0 + a_{a_0} \gamma^{a_0} + a_{a_1 a_2} \gamma^{a_1} \gamma^{a_2} + \cdots + 
a_{a_1 \cdots a_d} \gamma^{a_1}\cdots \gamma^{a_d} \,,
\label{tildegcliffordappendix}
\end{eqnarray}
where $|\psi_{0}>$ is a vacuum state, defined in Eq.~(\ref{graphherscal}) and 
$(-)^{n_B} $ is equal to $1$ for the term in the polynomial which has an even number of $\gamma^b$'s, 
and to $-1$ for the term with an odd number of  $\gamma^b$'s,
for any $d$, even or odd, and  $I$ is the unit element in the 
Clifford algebra. 

It follows from Eq.~(\ref{tildegcliffordappendix}) that the two kinds of the 
Clifford algebra objects are connected with
the left and the right multiplication of any Clifford algebra objects $B$
 (Eq.~(\ref{tildegcliffordappendix})).

%14.05 2016 at 21

The Clifford algebra objects $S^{ab}$ and $\tilde{S}^{ab}$ close the algebra of the Lorentz 
group % (Eq.~(\ref{properties1})) 
\begin{eqnarray}
\label{sabtildesab}
S^{ab}: &=& (i/4) (\gamma^a \gamma^b - \gamma^b \gamma^a)\,, \nonumber\\
\tilde{S}^{ab}: &=& (i/4) (\tilde{\gamma}^a \tilde{\gamma}^b 
- \tilde{\gamma}^b \tilde{\gamma}^a)\,,
\end{eqnarray}
%\nonumber\\
$ \{S^{ab}, \tilde{S}^{cd}\}_{-}= 0\,$, %\nonumber\\
$\{S^{ab},S^{cd}\}_{-} = $ $ i(\eta^{ad} S^{bc} + \eta^{bc} S^{ad} - \eta^{ac} S^{bd} - \eta^{bd} S^{ac})\,$,
%\nonumber\\
$\{\tilde{S}^{ab},\tilde{S}^{cd}\}_{-} $ $= i(\eta^{ad} \tilde{S}^{bc} + \eta^{bc} \tilde{S}^{ad} 
- \eta^{ac} \tilde{S}^{bd} - \eta^{bd} \tilde{S}^{ac})\,$.
%\end{eqnarray}
%

We assume  the ``Hermiticity'' property for $\gamma^a$'s  %and $\tilde{\gamma}^a$'s 
\begin{eqnarray}
\gamma^{a\dagger} = \eta^{aa} \gamma^a\,,%\quad \quad \tilde{\gamma}^{a\dagger} = \eta^{aa} \tilde{\gamma}^a\,,
\label{cliffher}
\end{eqnarray}
in order that 
$\gamma^a$ %and $\tilde{\gamma}^a$ 
are compatible with (\ref{gammatildegamma}) and formally unitary, 
i.e. $\gamma^{a \,\dagger} \,\gamma^a=I$. % and $\tilde{\gamma}^{a\,\dagger} \tilde{\gamma}^a=I$.
%We could make for \tilde{\gamma}^{a\dagger} =- \eta^{aa} \tilde{\gamma}^a\,, which would also be 
% acceptable as an antiunitary operator and everything would work. Use this for $d=(5+1)$.

One finds from Eq.~(\ref{cliffher}) that $(S^{ab})^{\dagger} = \eta^{aa} \eta^{bb}S^{ab}$.

Recognizing from Eq.(\ref{sabtildesab})  that the two Clifford algebra objects 
$S^{ab}, S^{cd}$ with all indices different commute, and equivalently for 
$\tilde{S}^{ab},\tilde{S}^{cd}$, we  select  the Cartan subalgebra of the algebra of the 
two groups, which  form  equivalent representations with respect to one another 
\begin{eqnarray}
S^{03}, S^{12}, S^{56}, \cdots, S^{d-1\; d}, \quad {\rm if } \quad d &=& 2n\ge 4,
\nonumber\\
S^{03}, S^{12}, \cdots, S^{d-2 \;d-1}, \quad {\rm if } \quad d &=& (2n +1) >4\,,
\nonumber\\
\tilde{S}^{03}, \tilde{S}^{12}, \tilde{S}^{56}, \cdots, \tilde{S}^{d-1\; d}, 
\quad {\rm if } \quad d &=& 2n\ge 4\,,
\nonumber\\
\tilde{S}^{03}, \tilde{S}^{12}, \cdots, \tilde{S}^{d-2 \;d-1}, 
\quad {\rm if } \quad d &=& (2n +1) >4\,.
\label{choicecartan}
\end{eqnarray}

The choice for  the Cartan subalgebra in $d <4$ is straightforward.
It is  useful  to define one of the Casimirs of the Lorentz group -  
the  handedness $\Gamma$ ($\{\Gamma, S^{ab}\}_- =0$) in any $d$ 
\begin{eqnarray}
\Gamma^{(d)} :&=&(i)^{d/2}\; \;\;\;\;\;\prod_a \quad (\sqrt{\eta^{aa}} \gamma^a), \quad {\rm if } \quad d = 2n, 
\nonumber\\
\Gamma^{(d)} :&=& (i)^{(d-1)/2}\; \prod_a \quad (\sqrt{\eta^{aa}} \gamma^a), \quad {\rm if } \quad d = 2n +1\,.
\label{hand}
\end{eqnarray}
One proceeds equivalently for $\tilde{\Gamma}^{(d)} $, substituting $\gamma^a$'s by $\tilde{\gamma}^a$'s.
We understand the product of $\gamma^a$'s in the ascending order with respect to 
the index $a$: $\gamma^0 \gamma^1\cdots \gamma^d$. 
It follows from Eq.(\ref{cliffher})
for any choice of the signature $\eta^{aa}$ that
$\Gamma^{\dagger}= \Gamma,\;
\Gamma^2 = I.$
We also find that for $d$ even the handedness  anticommutes with the Clifford algebra objects 
$\gamma^a$ ($\{\gamma^a, \Gamma \}_+ = 0$) , while for $d$ odd it commutes with  
$\gamma^a$ ($\{\gamma^a, \Gamma \}_- = 0$). 

To make the technique simple we introduce the graphic presentation 
as follows %(Eq.~(\ref{snmb:eigensab}))
\begin{eqnarray}
\stackrel{ab}{(k)}:&=& 
\frac{1}{2}(\gamma^a + \frac{\eta^{aa}}{ik} \gamma^b)\,,\quad \quad
\stackrel{ab}{[k]}:=
\frac{1}{2}(1+ \frac{i}{k} \gamma^a \gamma^b)\,,%\nonumber\\
%\stackrel{+}{\circ}:&=& \frac{1}{2} (1+\Gamma)\,,\quad \quad
%\stackrel{-}{\bullet}:= \frac{1}{2}(1-\Gamma),
\label{signature}
\end{eqnarray}
where $k^2 = \eta^{aa} \eta^{bb}$.
It follows then 
\begin{eqnarray}
\gamma^{a}&=& \stackrel{ab}{(k)} + \stackrel{ab}{(-k)}\,, \quad \quad 
\gamma^{b} = ik\eta^{aa}\,(\stackrel{ab}{(k)} - \stackrel{ab}{(-k)})\,,\nonumber\\
S^{ab}    &=& \frac{k}{2} (\stackrel{ab}{[k]}- \stackrel{ab}{[-k]})\,.
\label{signaturegamma}
\end{eqnarray}
One can easily check by taking into account the Clifford algebra relation 
(Eq.~(\ref{gammatildegamma})) and the
definition of $S^{ab}$ and $\tilde{S}^{ab}$ (Eq.~(\ref{sabtildesab}))
that the nilpotent $\stackrel{ab}{(k)}$ and the projector $\stackrel{ab}{[k]}$ are "eigenstates" of
$S^{ab}$ and $\tilde{S}^{ab}$ 
\begin{eqnarray}
        S^{ab}\, \stackrel{ab}{(k)}= \frac{1}{2}\,k\, \stackrel{ab}{(k)}\,,\quad \quad 
        S^{ab}\, \stackrel{ab}{[k]}= \frac{1}{2}\,k \,\stackrel{ab}{[k]}\,,\nonumber\\
\tilde{S}^{ab}\, \stackrel{ab}{(k)}= \frac{1}{2}\,k \,\stackrel{ab}{(k)}\,,\quad \quad 
\tilde{S}^{ab}\, \stackrel{ab}{[k]}=-\frac{1}{2}\,k \,\stackrel{ab}{[k]}\,,
\label{grapheigen}
\end{eqnarray}
which means that we get the same objects back multiplied by the constant $\frac{1}{2}k$ in the case 
of $S^{ab}$, while $\tilde{S}^{ab}$ multiply $\stackrel{ab}{(k)}$ by $k$ and $\stackrel{ab}{[k]}$ 
by $(-k)$ rather than $(k)$. 
%%%%%%%%%%%%
This also means that when 
$\stackrel{ab}{(k)}$ and $\stackrel{ab}{[k]}$ act from the left hand side on  a
vacuum state $|\psi_0\rangle$ the obtained states are the eigenvectors of $S^{ab}$.
We further recognize %~(Eq.~\ref{graphgammaaction},\ref{gammatilde}) 
that $\gamma^a$ 
transform  $\stackrel{ab}{(k)}$ into  $\stackrel{ab}{[-k]}$, never to $\stackrel{ab}{[k]}$, 
while $\tilde{\gamma}^a$ transform  $\stackrel{ab}{(k)}$ into $\stackrel{ab}{[k]}$, never to 
$\stackrel{ab}{[-k]}$ 
\begin{eqnarray}
%\label{snmb:graphgammatilgegammaaction}
&&\gamma^a \stackrel{ab}{(k)}= \eta^{aa}\stackrel{ab}{[-k]},\; 
\gamma^b \stackrel{ab}{(k)}= -ik \stackrel{ab}{[-k]}, \; 
\gamma^a \stackrel{ab}{[k]}= \stackrel{ab}{(-k)},\; 
\gamma^b \stackrel{ab}{[k]}= -ik \eta^{aa} \stackrel{ab}{(-k)}\,,\nonumber\\
&&\tilde{\gamma^a} \stackrel{ab}{(k)} = - i\eta^{aa}\stackrel{ab}{[k]},\;
\tilde{\gamma^b} \stackrel{ab}{(k)} =  - k \stackrel{ab}{[k]}, \;
\tilde{\gamma^a} \stackrel{ab}{[k]} =  \;\;i\stackrel{ab}{(k)},\; 
\tilde{\gamma^b} \stackrel{ab}{[k]} =  -k \eta^{aa} \stackrel{ab}{(k)}\,. 
\label{snmb:gammatildegamma}
\end{eqnarray}
From Eq.(\ref{snmb:gammatildegamma}) it follows
\begin{eqnarray}
\label{stildestrans}
S^{ac}\stackrel{ab}{(k)}\stackrel{cd}{(k)} &=& -\frac{i}{2} \eta^{aa} \eta^{cc} 
\stackrel{ab}{[-k]}\stackrel{cd}{[-k]}\,,\,\quad\quad
\tilde{S}^{ac}\stackrel{ab}{(k)}\stackrel{cd}{(k)} = \frac{i}{2} \eta^{aa} \eta^{cc} 
\stackrel{ab}{[k]}\stackrel{cd}{[k]}\,,\,\nonumber\\
S^{ac}\stackrel{ab}{[k]}\stackrel{cd}{[k]} &=& \frac{i}{2}  
\stackrel{ab}{(-k)}\stackrel{cd}{(-k)}\,,\,\quad\quad
\tilde{S}^{ac}\stackrel{ab}{[k]}\stackrel{cd}{[k]} = -\frac{i}{2}  
\stackrel{ab}{(k)}\stackrel{cd}{(k)}\,,\,\nonumber\\
S^{ac}\stackrel{ab}{(k)}\stackrel{cd}{[k]}  &=& -\frac{i}{2} \eta^{aa}  
\stackrel{ab}{[-k]}\stackrel{cd}{(-k)}\,,\,\quad\quad
\tilde{S}^{ac}\stackrel{ab}{(k)}\stackrel{cd}{[k]} = -\frac{i}{2} \eta^{aa}  
\stackrel{ab}{[k]}\stackrel{cd}{(k)}\,,\,\nonumber\\
S^{ac}\stackrel{ab}{[k]}\stackrel{cd}{(k)} &=& \frac{i}{2} \eta^{cc}  
\stackrel{ab}{(-k)}\stackrel{cd}{[-k]}\,,\,\quad\quad
\tilde{S}^{ac}\stackrel{ab}{[k]}\stackrel{cd}{(k)} = \frac{i}{2} \eta^{cc}  
\stackrel{ab}{(k)}\stackrel{cd}{[k]}\,. 
\end{eqnarray}
From Eq.~(\ref{stildestrans}) we conclude that $\tilde{S}^{ab}$ generate the 
equivalent representations with respect to $S^{ab}$ and opposite. 

Let us deduce some useful relations

\begin{eqnarray}
\stackrel{ab}{(k)}\stackrel{ab}{(k)}& =& 0\,, \quad \quad \stackrel{ab}{(k)}\stackrel{ab}{(-k)}
= \eta^{aa}  \stackrel{ab}{[k]}\,, \quad \stackrel{ab}{(-k)}\stackrel{ab}{(k)}=
\eta^{aa}   \stackrel{ab}{[-k]}\,,\quad
\stackrel{ab}{(-k)} \stackrel{ab}{(-k)} = 0\,, \nonumber\\
\stackrel{ab}{[k]}\stackrel{ab}{[k]}& =& \stackrel{ab}{[k]}\,, \quad \quad
\stackrel{ab}{[k]}\stackrel{ab}{[-k]}= 0\,, \;\;\quad \quad  \quad \stackrel{ab}{[-k]}\stackrel{ab}{[k]}=0\,,
 \;\;\quad \quad \quad \quad \stackrel{ab}{[-k]}\stackrel{ab}{[-k]} = \stackrel{ab}{[-k]}\,,
 \nonumber\\
\stackrel{ab}{(k)}\stackrel{ab}{[k]}& =& 0\,,\quad \quad \quad \stackrel{ab}{[k]}\stackrel{ab}{(k)}
=  \stackrel{ab}{(k)}\,, \quad \quad \quad \stackrel{ab}{(-k)}\stackrel{ab}{[k]}=
 \stackrel{ab}{(-k)}\,,\quad \quad \quad 
\stackrel{ab}{(-k)}\stackrel{ab}{[-k]} = 0\,,
\nonumber\\
\stackrel{ab}{(k)}\stackrel{ab}{[-k]}& =&  \stackrel{ab}{(k)}\,,
\quad \quad \stackrel{ab}{[k]}\stackrel{ab}{(-k)} =0,  \quad \quad 
\quad \stackrel{ab}{[-k]}\stackrel{ab}{(k)}= 0\,, \quad \quad \quad \quad
\stackrel{ab}{[-k]}\stackrel{ab}{(-k)} = \stackrel{ab}{(-k)}\,.\nonumber\\
\label{graphbinoms}
\end{eqnarray}
We recognize in Eq.~(\ref{graphbinoms}) 
the demonstration of the nilpotent and the projector character of the Clifford algebra objects 
$\stackrel{ab}{(k)}$ and $\stackrel{ab}{[k]}$, respectively. 
Defining
\begin{eqnarray}
\stackrel{ab}{\tilde{(\pm i)}} = 
\frac{1}{2} \, (\tilde{\gamma}^a \mp \tilde{\gamma}^b)\,, \quad
\stackrel{ab}{\tilde{(\pm 1)}} = 
\frac{1}{2} \, (\tilde{\gamma}^a \pm i\tilde{\gamma}^b)\,, 
%\stackrel{ab}{\tilde{[\pm i]}} = \frac{1}{2} (1 \pm \tilde{\gamma}^a \tilde{\gamma}^b), \quad
%\stackrel{ab}{\tilde{[\pm 1]}} = \frac{1}{2} (1 \pm i \tilde{\gamma}^a \tilde{\gamma}^b). \nonumber
\label{deftildefun}
\end{eqnarray}
one recognizes that
\begin{eqnarray}
\stackrel{ab}{\tilde{( k)}} \, \stackrel{ab}{(k)}& =& 0\,, 
\quad \;
\stackrel{ab}{\tilde{(-k)}} \, \stackrel{ab}{(k)} = -i \eta^{aa}\,  \stackrel{ab}{[k]}\,,
\quad\;
\stackrel{ab}{\tilde{( k)}} \, \stackrel{ab}{[k]} = i\, \stackrel{ab}{(k)}\,,
\quad\;
\stackrel{ab}{\tilde{( k)}}\, \stackrel{ab}{[-k]} = 0\,.
\label{graphbinomsfamilies}
\end{eqnarray}
Recognizing that
\begin{eqnarray}
\stackrel{ab}{(k)}^{\dagger}=\eta^{aa}\stackrel{ab}{(-k)}\,,\quad
\stackrel{ab}{[k]}^{\dagger}= \stackrel{ab}{[k]}\,,
\label{graphherstr}
\end{eqnarray}
we define a vacuum state $|\psi_0>$ so that one finds
\begin{eqnarray}
< \;\stackrel{ab}{(k)}^{\dagger}
 \stackrel{ab}{(k)}\; > = 1\,, \nonumber\\
< \;\stackrel{ab}{[k]}^{\dagger}
 \stackrel{ab}{[k]}\; > = 1\,.
\label{graphherscal}
\end{eqnarray}

Taking into account the above equations it is easy to find a Weyl spinor irreducible representation
for $d$-dimensional space, with $d$ even or odd.

For $d$ even we simply make a starting state as a product of $d/2$, let us say, only nilpotents 
$\stackrel{ab}{(k)}$, one for each $S^{ab}$ of the Cartan subalgebra  elements (Eq.(\ref{choicecartan})),  
applying it on an (unimportant) vacuum state. 
For $d$ odd the basic states are products
of $(d-1)/2$ nilpotents and a factor $(1\pm \Gamma)$.  
Then the generators $S^{ab}$, which do not belong 
to the Cartan subalgebra, being applied on the starting state from the left, 
 generate all the members of one
Weyl spinor.  
\begin{eqnarray}
\stackrel{0d}{(k_{0d})} \stackrel{12}{(k_{12})} \stackrel{35}{(k_{35})}\cdots \stackrel{d-1\;d-2}{(k_{d-1\;d-2})}
|\psi_0 \,>\nonumber\\
\stackrel{0d}{[-k_{0d}]} \stackrel{12}{[-k_{12}]} \stackrel{35}{(k_{35})}\cdots \stackrel{d-1\;d-2}{(k_{d-1\;d-2})}
|\psi_0 \,>\nonumber\\
\stackrel{0d}{[-k_{0d}]} \stackrel{12}{(k_{12})} \stackrel{35}{[-k_{35}]}\cdots \stackrel{d-1\;d-2}{(k_{d-1\;d-2})}
|\psi_0 \,>\nonumber\\
\vdots \nonumber\\
\stackrel{0d}{[-k_{0d}]} \stackrel{12}{(k_{12})} \stackrel{35}{(k_{35})}\cdots \stackrel{d-1\;d-2}{[-k_{d-1\;d-2}]}
|\psi_0 \,>\nonumber\\
\stackrel{od}{(k_{0d})} \stackrel{12}{[-k_{12}]} \stackrel{35}{[-k_{35}]}\cdots \stackrel{d-1\;d-2}{(k_{d-1\;d-2})}
|\psi_0\,> \nonumber\\
\vdots 
\label{graphicd}
\end{eqnarray}
All the states have the same handedness $\Gamma $, since $\{ \Gamma, S^{ab}\}_{-} = 0$. 
States, belonging to one multiplet  with respect to the group $SO(q,d-q)$, that is to one
irreducible representation of spinors (one Weyl spinor), can have any phase. We made a choice
of the simplest one, taking all  phases equal to one.

The above graphic representation demonstrates that for $d$ even 
all the states of one irreducible Weyl representation of a definite handedness follow from a starting state, 
which is, for example, a product of nilpotents $\stackrel{ab}{(k_{ab})}$, by transforming all possible pairs
of $\stackrel{ab}{(k_{ab})} \stackrel{mn}{(k_{mn})}$ into $\stackrel{ab}{[-k_{ab}]} \stackrel{mn}{[-k_{mn}]}$.
There are $S^{am}, S^{an}, S^{bm}, S^{bn}$, which do this.
The procedure gives $2^{(d/2-1)}$ states. A Clifford algebra object $\gamma^a$ being applied from the left hand side,
transforms  a 
Weyl spinor of one handedness into a Weyl spinor of the opposite handedness. Both Weyl spinors form a Dirac 
spinor.

%For $d$ odd a Weyl spinor has besides a product of $(d-1)/2$ nilpotents or projectors also either the
%factor $\stackrel{+}{\circ}:= \frac{1}{2} (1+\Gamma)$ or the factor
%$\stackrel{-}{\bullet}:= \frac{1}{2}(1-\Gamma)$.  
%As in the case of $d$ even, all the states of one irreducible 
%Weyl representation of a definite handedness follow from a starting state, 
%which is, for example, a product of $(1 + \Gamma)$ and $(d-1)/2$ nilpotents $\stackrel{ab}{(k_{ab})}$, by 
%transforming all possible pairs
%of $\stackrel{ab}{(k_{ab})} \stackrel{mn}{(k_{mn})}$ into $\stackrel{ab}{[-k_{ab}]} \stackrel{mn}{[-k_{mn}]}$.
%But $\gamma^a$'s, being applied from the left hand side, do not change the handedness of the Weyl spinor, 
%since $\{ \Gamma,
%\gamma^a \}_- =0$ for $d$ odd. 
%A Dirac and a Weyl spinor are for $d$ odd identical and a ''family'' 
%has accordingly $2^{(d-1)/2}$ members of basic states of a definite handedness.

We shall speak about left handedness when $\Gamma= -1$ and about right
handedness when $\Gamma =1$ for either $d$ even or odd.

While $S^{ab}$ which do not belong to the Cartan subalgebra (Eq.~(\ref{choicecartan})) generate 
all the states of one representation,  $\tilde{S}^{ab}$ which do not belong to the 
Cartan subalgebra (Eq.~(\ref{choicecartan}))  generate the states of $2^{d/2-1}$ equivalent representations.

Making a choice of the Cartan subalgebra set (Eq.~(\ref{choicecartan})) 
of the algebra $S^{ab}$ and 
$\tilde{S}^{ab}$  
%
%\begin{eqnarray}
$S^{03}, S^{12}, S^{56}, S^{78}, S^{9 \;10}, S^{11\;12}, S^{13\; 14}\,$, %\nonumber\\
$\tilde{S}^{03}, \tilde{S}^{12}, \tilde{S}^{56}, \tilde{S}^{78}, \tilde{S}^{9 \;10}, 
\tilde{S}^{11\;12}, \tilde{S}^{13\; 14}\,$,
%\label{cartan}
%\end{eqnarray}
%
a left handed ($\Gamma^{(13,1)} =-1$) eigenstate of all the members of the 
Cartan  subalgebra, representing a weak chargeless  $u_{R}$-quark with spin up, hyper charge ($2/3$) 
and  colour ($1/2\,,1/(2\sqrt{3})$), for example, can be written as %(Eq.(\ref{cartan})) 
\begin{eqnarray}
&& \stackrel{03}{(+i)}\stackrel{12}{(+)}|\stackrel{56}{(+)}\stackrel{78}{(+)}
||\stackrel{9 \;10}{(+)}\stackrel{11\;12}{(-)}\stackrel{13\;14}{(-)} |\psi_{0} \rangle = \nonumber\\
&&\frac{1}{2^7} 
(\gamma^0 -\gamma^3)(\gamma^1 +i \gamma^2)| (\gamma^5 + i\gamma^6)(\gamma^7 +i \gamma^8)||
\nonumber\\
&& (\gamma^9 +i\gamma^{10})(\gamma^{11} -i \gamma^{12})(\gamma^{13}-i\gamma^{14})
|\psi_{0} \rangle \,.
\label{start}
\end{eqnarray}
This state is an eigenstate of all $S^{ab}$ and $\tilde{S}^{ab}$ which are members of the Cartan 
subalgebra (Eq.~(\ref{choicecartan})). 

The operators $ \tilde{S}^{ab}$, which do not belong to the Cartan subalgebra (Eq.~(\ref{choicecartan})),  
generate families from the starting $u_R$ quark, transforming the $u_R$ quark from Eq.~(\ref{start}) 
to the $u_R$ of another family,  keeping all of the properties with respect to $S^{ab}$ unchanged.
In particular, $\tilde{S}^{01}$ applied on a right handed $u_R$-quark 
%, weak chargeless,  with spin up,hypercharge ($2/3$) and the colour charge ($1/2\,,1/(2\sqrt{3})$) 
from Eq.~(\ref{start}) generates a 
state which is again  a right handed $u_{R}$-quark,  weak chargeless,  with spin up,
hyper charge ($2/3$)
and the colour charge ($1/2\,,1/(2\sqrt{3})$)
\begin{eqnarray}
\tilde{S}^{01}\;
\stackrel{03}{(+i)}\stackrel{12}{(+)}| \stackrel{56}{(+)} \stackrel{78}{(+)}||
\stackrel{9 10}{(+)} \stackrel{11 12}{(-)} \stackrel{13 14}{(-)}= -\frac{i}{2}\,
&&\stackrel{03}{[\,+i]} \stackrel{12}{[\,+\,]}| \stackrel{56}{(+)} \stackrel{78}{(+)}||
\stackrel{9 10}{(+)} \stackrel{11 12}{(-)} \stackrel{13 14}{(-)}\,.\nonumber\\
\label{tildesabfam}
\end{eqnarray}

Below some useful relations~\cite{pikanorma} are presented 
\begin{eqnarray}
\label{plusminus}
N^{\pm}_{+}         &=& N^{1}_{+} \pm i \,N^{2}_{+} = 
 - \stackrel{03}{(\mp i)} \stackrel{12}{(\pm )}\,, \quad N^{\pm}_{-}= N^{1}_{-} \pm i\,N^{2}_{-} = 
  \stackrel{03}{(\pm i)} \stackrel{12}{(\pm )}\,,\nonumber\\
\tilde{N}^{\pm}_{+} &=& - \stackrel{03}{\tilde{(\mp i)}} \stackrel{12}{\tilde{(\pm )}}\,, \quad 
\tilde{N}^{\pm}_{-}= %\tilde{N}^{1}_{-} \pm i\,\tilde{N}^{2}_{-} = 
  \stackrel{03} {\tilde{(\pm i)}} \stackrel{12} {\tilde{(\pm )}}\,,\nonumber\\ 
\tau^{1\pm}         &=& (\mp)\, \stackrel{56}{(\pm )} \stackrel{78}{(\mp )} \,, \quad   
\tau^{2\mp}=            (\mp)\, \stackrel{56}{(\mp )} \stackrel{78}{(\mp )} \,,\nonumber\\ 
\tilde{\tau}^{1\pm} &=& (\mp)\, \stackrel{56}{\tilde{(\pm )}} \stackrel{78}{\tilde{(\mp )}}\,,\quad   
\tilde{\tau}^{2\mp}= (\mp)\, \stackrel{56}{\tilde{(\mp )}} \stackrel{78}{\tilde{(\mp )}}\,.
\end{eqnarray}
%

%%%%%%%%%%%%%%%%%%% MOVE below somewhere!
%One Weyl representation of $SO(13+1)$  contains left handed weak charged and 
%the second $SU(2)$ chargeless coloured quarks and colourless leptons and right handed weak chargeless
%and the second $SU(2)$ charged quarks and leptons (electrons and neutrinos). It carries also the family
%quantum numbers, not mentioned in this table. The table is taken from Ref.~\cite{discretesym}.
%%%%%%%%%%%%%%%%%%%%%

%Text...
\end{appendix}

%\bibliographystyle{ws-rv-van}
%\bibliography{ws-rv-sample}

\begin{thebibliography}{9}
\bibitem{NPLB} N.S. Manko\v c Bor\v stnik, "Do we have the explanation for
the Higgs and Yukawa couplings of the {\em standard model}", [arxiv:1212.3184,
arxiv:1011.5765].
\end{thebibliography}
%\printindex[aindx]                % to print author index
\printindex
\end{document}